\documentclass[12pt]{article}
\usepackage[a4paper]{geometry}
\usepackage{amsfonts,amssymb,amsmath,enumerate}
\usepackage{color}
\usepackage{theorem,cite}
\usepackage{indentfirst}
\usepackage{textcomp}
\usepackage{mathdots}
\usepackage{bm}
\usepackage{hyperref}
\newcommand{\prf}{\textsl{Proof:}\ }

\theorembodyfont{\rmfamily}

\newcommand{\mb}[1]{\hspace{2.1ex}\mbox{#1}\hspace{2.1ex}}
\newcommand{\nonu}{\nonumber\\ }
 \makeatletter
 \@addtoreset{equation}{section}
 \makeatother

\newcommand\ket[1]{{|#1\rangle}}

\newcommand\bbra[1]{{\langle\!\langle#1|}}
\newcommand\kket[1]{{|#1\rangle\!\rangle}}

\newcommand{\llangle}{\langle\!\langle}
\newcommand{\rrangle}{\rangle\!\rangle}

\def\steady{|{\cal S}\rangle}

\newcommand\opd{\mathbf{d}}
\newcommand\ope{\mathbf{e}}
\newcommand\opA{\mathbf{A}}
\newcommand\opid{\mathbf{1}}

\newcommand\genfracz[2]{\genfrac{}{}{0pt}{}{#1}{#2}}

\def\cA{{\cal A}}    \def\cB{{\cal B}}    \def\cC{{\cal C}}
\def\cD{{\cal D}}    \def\cE{{\cal E}}

\def\cM{{\cal M}}        
        \def\cR{{\cal R}}
\def\cS{{\cal S}}        
\def\cV{{\cal V}}    \def\cW{{\cal W}}     
    \def\cZ{{\cal Z}}    

\def\fB{{\mathfrak B}}

\def\fX{{\mathfrak X}}

\def\fb{{\mathfrak b}}

\def\fz{{\mathfrak z}}

\newcommand{\CC}{{\mathbb C}}

\begin{document}
\rightline{LAPTh-059/17}
\rightline{December 2017}
\vfill

\begin{center}

 {\LARGE  {\sffamily Matrix product solution to multi-species ASEP\\ with open boundaries} }\\[1cm]

\vspace{10mm}
  
{\Large 
 C. Finn$^{ab}$\footnote{cfinn@unimelb.edu.au},
 E. Ragoucy$^{a}$\footnote{eric.ragoucy@lapth.cnrs.fr}
 and M. Vanicat$^{ac}$\footnote{matthieu.vanicat@fmf.uni-lj.si}}\\[.41cm] 
$^{a}$ Laboratoire de Physique Th{\'e}orique LAPTh,
 CNRS and Universit{\'e} Savoie Mont Blanc.\\
   9 chemin de Bellevue, BP 110, F-74941  Annecy-le-Vieux Cedex, 
France. 
\\[.42cm]
$^{b}$ ARC Centre of Excellence for Mathematical and Statistical Frontiers (ACEMS), \\
 School of Mathematics and Statistics, University of Melbourne, Victoria 3010, Australia.
\\[.42cm]
$^{c}$ Faculty of Mathematics and Physics, University of Ljubljana,\\
 Jadranska 19, SI-1000 Ljubljana, Slovenia.
\end{center}
\vfill

\begin{abstract}
We study a class of multi-species ASEP with open boundaries. The boundaries are chosen in such a way that all species of particles 
interact non-trivially with the boundaries and are present in the stationary state.
We give the exact expression of the stationary state in a matrix product form and we compute its normalisation.
Densities and currents for the different species are then computed in term of this normalisation.
\end{abstract}

\vfill
\newpage
\pagestyle{plain}
\section*{Introduction} 

The Asymmetric Simple Exclusion Process (ASEP) \cite{Liggett85,Spitzer70} has been widely studied in the last decades both in 
the physics and mathematics literature \cite{ChouMZ11,Derrida98}.
From a physical point of view, it has been first introduced in the context of biology and has found also later applications to 
traffic flow. It appears as one of the simplest models to possess the main physical properties of non-equilibrium systems, 
such as boundary induced phase transitions \cite{Krug91,SchuetzD93} or shock waves \cite{Ferrari91,FerrariKS91}.
From a mathematical perspective, the ASEP (with periodic or open boundaries conditions) enjoys the property of being integrable and displays
a rich structure.
In the open case the stationary state has been exactly computed in a matrix product form \cite{DerridaEHP93,Sandow94} 
and connections with combinatorics
and symmetric polynomials has been revealed \cite{UchiyamaSW04,FinnV17}. The study of the ASEP on the infinite line has also attracted a lot of attention because
of the connection with the Kardar-Parisi-Zhang universality class and has in this way contributed to the emergence of the field of 
\textit{integrable probability} \cite{BorodinFPS07,TracyW09}.

The huge success of the single species ASEP strongly motivated the definition and study of multi-species generalisations.
The N-species ASEP with periodic boudary conditions and homogeneous hopping rates still enjoys the integrable structure
and the steady state has been computed exactly in a matrix product form \cite{AAMP1,AAMP2,CantiniDGW15,EvansFM09,KunibaMO15,KunibaMO16}. 
Integrable deformations with inhomogeneous hopping rates 
between the different species have also been considered \cite{arita}. A lot of effort has also been put into the open boundaries case:
some 2-species generalisations of the ASEP with open boundaries have been studied, with semi-permeable boundaries
\cite{ALS,Uchiyama08} and with integrable boundaries \cite{caley,CrampeEMRV16,CrampeMRV15}. 
Specific N-species cases with open boundaries have been studied in \cite{Mandelshtam15} 
or with reflexive boundaries in \cite{Ari} and a large class of integrable boundaries have been provided in \cite{CrampeFRV16}.
Some connections of the stationary state with orthogonal polynomials have been revealed \cite{Cantini15,CantiniGDGW16}. 
The phase diagram of such multi-species ASEP with open boundaries has been exactly computed in some specific cases \cite{AyyerR16}.

In this work we propose to study a multi-species ASEP with open boundaries. The injection and extraction rates at the boundaries describe a 
non-trivial interaction of the different species with the reservoirs and ensure that all species of particle are present with non-vanishing probability
in the stationary state. The left (respectively right) boundary conditions depend on $N$ or $N-1$ (respectively $N+1$ or $N$) independent parameters,
depending on the parity of the number $N$ of different species in the model. These boundary conditions do not fall in the class of 
integrable boundaries presented in \cite{CrampeFRV16}.
Nevertheless they reduce to integrable boundaries introduced in \cite{CrampeFRV16} when the parameters are specified to some specific values.
For general parameters, we provide an exact expression of the stationary state in a matrix product form. The matrices are built as tensor products 
of the $\ope$, $\opd$ and $\opA$ operators introduced to solve the single species ASEP with open boundaries and the 2-species 
ASEP on the ring.
This simple algebraic structure allows us to compute the normalisation of the stationary state and to express the particle currents and densities.

In section \ref{sec:Markovian} we introduce the formalism we use to define the Markovian dynamics of our model. We define the dynamics of the 
different species in the bulk of the lattice and we write down the expression of the Markov matrix. Section \ref{sec:2ASEP} is devoted to the 
study of the simplest case: the 2-species model. The matrix product construction of the stationary state is presented. This 2-species model and 
its matrix product solution will serve as the building block of the N-species model, and of the construction of the corresponding stationary state in 
a matrix product form. We comment also on the symmetric and totally asymmetric limits. In section \ref{sec:3ASEP} we briefly present, 
for pedagogical reasons, the 3-species case, to introduce the tensor structure of the matrix product solution and to show how 
the normalisation and the particle currents and densities can be computed thanks to this algebraic structure.
We come finally in section \ref{sec:NASEP} to the N-species case. We define precisely the boundary conditions of the model and we provide 
an exact expression of the stationary state in a matrix product form. This allows us to derive an expression of the normalisation and of the mean
particle currents and densities.

\section{Preliminaries \label{sec:Markovian}}

In the present work, we are interested in a multi-species ASEP with open boundary conditions. We consider $N$ species of 
particles respectively labeled by $1$, $2$, ..., $N$. The holes (vacancies) on the lattice are denoted by $0$. 
We attach to each site $i$ of the lattice an occupation variable $\tau_i \in \{0,1,...,N\}$, which specifies the species of the particle lying on the 
site if it is occupied or which is equal to $0$ if the site is empty. A configuration of particles on the whole lattice is thus concisely 
described by a $L$-uplet $\bm{\tau}=(\tau_1,\tau_2,...,\tau_L)$.

The stochastic dynamics of the system is Markovian: during an infinitesimal interval of time $\delta t$ a local configuration $\tau_i$,$\tau_{i+1}$
on two adjacent sites $i$ and $i+1$ in the bulk is swapped with probability $q\times \delta t$ (respectively with probability $\delta t$) if 
$\tau_i < \tau_{i+1}$ (respectively if $\tau_i > \tau_{i+1}$). In this paper we consider $q \leq 1$, so that the different species can be ordered
regarding the dynamics: species $N$ can be thought as the fastest species whereas species $1$ can be interpreted as the slowest one.
On the boundaries the local configurations $\tau_1$ on the first site and $\tau_L$
on the last site can be modified with probability rates depending on the species of particle involved. These rates will be specified below.

This Markovian dynamics can be easily encoded using a transition matrix. We must first specify a basis: 
we associate to each configuration $\bm{\tau}$ a basis vector $\ket{\bm{\tau}}$ constructed as a tensor product
\begin{equation}
 \ket{\bm{\tau}} = \ket{\tau_1} \otimes \ket{\tau_2} \otimes \cdots \otimes \ket{\tau_L},
\end{equation}
where the elementary vectors $\ket{\tau} \in \CC^{N+1}$ are defined by 
$\ket{\tau}=(\underbrace{0,\dots,0}_{\tau},1,\underbrace{0,\dots,0}_{N-\tau})^t$.

This allows us to encompass the probabilities $P_t(\bm{\tau})$ to observe the system in configuration $\bm{\tau}$ at time $t$ in a single vector
\begin{equation}
 \ket{P_t} = \sum_{\bm{\tau}} P_t(\bm{\tau}) \ket{\bm{\tau}}.
\end{equation}
The time evolution of this probability vector obeys the master equation 
\begin{equation}
 \frac{d\ket{P_t}}{dt} = M \ket{P_t},
\end{equation}
where $M$ is the Markov matrix of the model. Because the stochastic dynamics involves only local moves on the lattice 
(exchange of two neighbor particles for instance), the transition matrix can be decomposed as the sum of operators acting locally on 
the lattice
\begin{equation}
 M= B_1 + \sum_{k=1}^{L-1} m_{k,k+1} + \overline{B}_L.
\end{equation}
The indices denote the components of the tensor space $\left(\CC^{N+1}\right)^{\otimes L}$ on which the operators act non-trivially. More precisely
we have 
\begin{equation}
 B_1 = B \otimes \opid \otimes \cdots \otimes \opid, \qquad \overline{B}_L = \opid \otimes \cdots \otimes \opid \otimes \overline{B}
\end{equation}
where $B$ and $\overline{B}$ are $(N+1) \times (N+1)$ matrices encoding respectively the dynamics with the left and right reservoir (an explicit 
expression of these matrices will be provided below) and
\begin{equation}
 m_{k,k+1} = \underbrace{\opid \otimes \cdots \otimes \opid}_{k-1} \otimes\, m \otimes \underbrace{\opid \otimes \cdots \otimes \opid}_{L-k-1},
\end{equation}
where $m$ is a $(N+1)^2 \times (N+1)^2$ matrix encoding the local dynamics on two adjacent sites in the bulk. Its explicit expression is given by
\begin{equation}
 m = \sum_{0 \leq i < j \leq N} \Big(E_{ij} \otimes E_{ji} + q \, E_{ji} \otimes E_{ij} - E_{jj} \otimes E_{ii} - q \, E_{ii} \otimes E_{jj} \Big).
\end{equation}

We will be interested in the steady state of the model, that is in a vector $\steady=\sum_{\bm{\tau}} \cS(\bm{\tau}) \ket{\bm{\tau}}$
in the kernel of the Markov matrix 
\begin{equation}
 M\steady =0.
\end{equation}

\section{A 2-species model \label{sec:2ASEP}}

We briefly recall the definition of a 2-species ASEP with integrable open boundaries.
It has been studied in detail in \cite{caley}, where the matrix product construction of the stationary state has been provided and the exact
phase diagram computed. We recall here the matrix product solution and some results about the computation of the normalisation and of 
the mean particle currents and densities.

We introduce this model here because it will play the role of a building block to construct more complicated multi-species models
presented in sections \ref{sec:3ASEP} and \ref{sec:NASEP}.

\subsection{Presentation of the model}
The model is a generalisation of the usual ASEP model. It contains two different species of particles (labelled 1 and 2) together with vacancies or empty sites (labelled 0). 
The local stochastic rules of the model are summarized in the following table. The probability rates corresponding to each transitions are 
written above the arrows.
\begin{eqnarray}
\begin{array}{|c|c|c|} 
\hline 
\mbox{Left} & \mbox{Bulk} & \mbox{Right} \\ \hline 
\hspace{0.5cm} 0\, \xrightarrow{\ \gamma\ } \,1\hspace{0.5cm}&\hspace{0.5cm} 10\, \xrightarrow{\ 1\ } \,01\hspace{0.5cm},\hspace{0.5cm} 01\, \xrightarrow{\ q\ }\, 10\hspace{0.5cm}&\hspace{0.5cm} 0\, \xrightarrow{\ \delta\ } \,2\hspace{0.5cm}\\
\hspace{0.5cm} 0\, \xrightarrow{\ \alpha\ } \,2\hspace{0.5cm}&\hspace{0.5cm} 20\, \xrightarrow{\ 1\ }\, 02\hspace{0.5cm},\hspace{0.5cm} 02\, \xrightarrow{\ q\ }\, 20\hspace{0.5cm}&\hspace{0.5cm} 2\, \xrightarrow{\ \beta\ } \,0\hspace{0.5cm}\\
\hspace{0.5cm} 1\, \xrightarrow{\ \alpha\ } \,2\hspace{0.5cm}&\hspace{0.5cm} 21\, \xrightarrow{\ 1\ } \,12\hspace{0.5cm},\hspace{0.5cm} 12\, \xrightarrow{\ q\ }\, 21\hspace{0.5cm}& \\
\hspace{0.5cm} 2\, \xrightarrow{\ \tilde\gamma\ } \,1\hspace{0.5cm}& & \\
\hline
 \end{array}
\end{eqnarray}
In words, the left boundary is \textit{permeable}: particles of species $1$ and $2$ can be both injected and absorbed by the left reservoir with 
given probability rates.
Conversely the right boundary is \textit{semi-permeable}: only particles of species $2$ can be injected and absorbed by the right reservoir, the 
right boundary behaves as a reflexive boundary for the particles of species $1$.
We will thus observe a vanishing mean current of particles of species $1$ in the stationary state. The system is nevertheless driven out-of-equilibrium
because it displays a non-vanishing current of particles of species $2$ in the stationary state.

As already mentioned, the stochastic rules introduced above can be efficiently recast in a matrix form, 
using a Markov matrix which takes the explicit expression 
\begin{equation}
 \cM = \cB_1 + \sum_{k=1}^{L-1} m_{k,k+1} + \overline{\cB}_L
\end{equation}
with the local bulk Markov matrix
\begin{equation}
 m = \begin{pmatrix}
      0 & 0 & 0 & 0 & 0 & 0 & 0 & 0 & 0 \\
      0 & -q & 0 & 1 & 0 & 0 & 0 & 0 & 0 \\
      0 & 0 & -q & 0 & 0 & 0 & 1 & 0 & 0 \\
      0 & q & 0 & -1 & 0 & 0 & 0 & 0 & 0 \\
      0 & 0 & 0 & 0 & 0 & 0 & 0 & 0 & 0 \\
      0 & 0 & 0 & 0 & 0 & -q & 0 & 1 & 0 \\
      0 & 0 & q & 0 & 0 & 0 & -1 & 0 & 0 \\
      0 & 0 & 0 & 0 & 0 & q & 0 & -1 & 0 \\
      0 & 0 & 0 & 0 & 0 & 0 & 0 & 0 & 0
     \end{pmatrix}
\end{equation}
and the boundary matrices
\begin{equation}
 \cB = \begin{pmatrix}
      -\alpha-\gamma & 0 & 0 \\
      \gamma & -\alpha & \tilde\gamma \\
      \alpha & \alpha & -\tilde\gamma
     \end{pmatrix} \quad \mbox{and} \quad
 \overline{\cB} = \begin{pmatrix}
                 -\delta & 0 & \beta \\
                 0 & 0 & 0 \\
                 \delta & 0 & -\beta 
                \end{pmatrix},
\end{equation}
where
\begin{equation}
 \tilde\gamma = \frac{\gamma(\alpha+\gamma+q-1)}{\alpha+\gamma}.
\end{equation}
We recall that the very particular shape of the boundary matrices $\cB$ and $\overline{\cB}$ and the value of the parameter $\tilde\gamma$ are
dictated by integrability. These two matrices are indeed picked up from the classification of integrable Markovian boundary conditions 
performed in \cite{CrampeFRV16} by solving the reflection equation (which is the boundary analogue of the Yang-Baxter equation).

\subsection{Matrix product solution}

\subsubsection{Algebraic relations}

Following \cite{caley}, the stationary state $\steady$ of the model, satisfying $\cM \steady =0$, is given in a matrix product form
\begin{equation} \label{eq:2ASEP_Matrix_Ansatz}
 \steady = \frac{1}{\cZ_L} \bbra{\cW} \begin{pmatrix}
                                       \cE \\ \cA \\ \cD
                                      \end{pmatrix} \otimes
                                      \begin{pmatrix}
                                       \cE \\ \cA \\ \cD
                                      \end{pmatrix} \otimes \dots \otimes
                                      \begin{pmatrix}
                                       \cE \\ \cA \\ \cD
                                      \end{pmatrix} \kket{\cV},
\end{equation}
where the normalisation is expressed as $\cZ_L = \bbra{\cW}(\cE + \cA + \cD)^L \kket{\cV}$.

The matrices satisfy the following algebraic relations
\begin{equation}\label{eq:comEAD}
\begin{aligned}
 & \cD\cE-q\cE\cD = \cE+\cD, \\
 & \cA\cE-q\cE\cA = \cA, \\
 & \cD\cA-q\cA\cD = \cA,
\end{aligned}
\end{equation}
 while the boundary vectors have to satisfy the relations below, which impose a specific representation for the $(\cA,\cD,\cE)$ algebra:
 \begin{equation}\label{eq:boundAED}
  \begin{cases}
   \bbra{\cW} \cE = \frac{1}{\alpha+\gamma}\bbra{\cW} \\
   \bbra{\cW} \left(\gamma \cE - \alpha \cA + \tilde\gamma \cD \right) = 0
  \end{cases}
 \quad \mbox{and} \quad \left(\delta \cE - \beta \cD \right)\kket{\cV} = -\kket{\cV}.
 \end{equation}
 
These equations are equivalent to the telescopic relations\footnote{It can be proven by direct computation, 
writing the telescopic relations in components.}
\begin{equation}
 m \begin{pmatrix}
   \cE \\ \cA \\ \cD
   \end{pmatrix} \otimes
   \begin{pmatrix}
   \cE \\ \cA \\ \cD
   \end{pmatrix} = 
   \begin{pmatrix}
   \cE \\ \cA \\ \cD
   \end{pmatrix} \otimes
   \begin{pmatrix}
   -1 \\ 0 \\ 1
   \end{pmatrix}-
   \begin{pmatrix}
   -1 \\ 0 \\ 1
   \end{pmatrix} \otimes
   \begin{pmatrix}
   \cE \\ \cA \\ \cD
   \end{pmatrix}
\end{equation}
and
\begin{equation}
 \bbra{\cW}\cB \begin{pmatrix}
               \cE \\ \cA \\ \cD
               \end{pmatrix} =
 \bbra{\cW} \begin{pmatrix}
            -1 \\ 0 \\ 1
            \end{pmatrix}, \qquad 
 \overline{\cB} \begin{pmatrix}
                \cE \\ \cA \\ \cD
                \end{pmatrix} \kket{\cV} =   
 -\begin{pmatrix}
  -1 \\ 0 \\ 1
  \end{pmatrix} \kket{\cV}
\end{equation}
which ensure that \eqref{eq:2ASEP_Matrix_Ansatz} provides a correct expression of the steady state.
We refer the interested reader to for instance \cite{CrampeRV14} for a proof.

 \subsubsection{Representation}
 An explicit representation of the operators $\cE$, $\cA$ and $\cD$ is provided in terms of the $q$-deformed oscillator algebra generated by 
 $\ope$, $\opd$, $\opA$ \cite{caley}
 \begin{equation} \label{def:alg_eAd}
  \begin{aligned}
   & \opd \ope -q\ope \opd = 1-q, \\
   & \opd \opA = q \opA \opd, \\
   & \opA \ope = q \ope \opA.
  \end{aligned}
 \end{equation}
 The expression is given by
 \begin{equation}\label{def:EAD}
\begin{aligned}
 & \cE = \frac{1}{1-q}(\opid+\ope), \\
 & \cA =\frac{r_0}{1-q}\opA, \quad\mbox{with}\quad r_0=\frac{\gamma}{\alpha}\\
 & \cD =\frac{1}{1-q}(\opid+\opd),
\end{aligned}
\end{equation}
The operators $\ope$, $\opd$, $\opA$ are matrices acting on the Fock space spanned by the basis $\{ \kket{n} \}_{n=0}^\infty$
\begin{equation*}
    \opd = \sum_{n=1}^\infty \sqrt{1 - q^n}\, \kket{n-1}\bbra{n},
    \qquad
    \ope = \sum_{n=0}^\infty \sqrt{1 - q^{n+1}}\, \kket{n+1}\bbra{n},
\end{equation*}
and
\begin{equation*}
    \opA = \sum_{n=0}^\infty q^n \kket{n}\bbra{n} = \frac{1}{1-q}(\opd\ope-\ope\opd)=1-\ope\opd.
\end{equation*}

It is easy to verify that, once the realisation \eqref{def:EAD} is chosen, the boundary conditions \eqref{eq:boundAED} reduce to
\begin{equation} \label{eq:boundaed}
 \bbra{\cW} \ope = c \bbra{\cW}, \quad \mbox{and} \quad (\opd+bd\ope)\kket{\cV} = (b+d)\kket{\cV}
\end{equation}
with 
\begin{equation}
 c = \frac{1-q-\alpha-\gamma}{\alpha+\gamma}
\end{equation}
and 
\begin{equation}
\begin{aligned}
 & b= \frac{1-q+\delta-\beta+\sqrt{(1-q+\delta-\beta)^2+4\beta\delta}}{2\beta}, \\ 
 & d= \frac{1-q+\delta-\beta-\sqrt{(1-q+\delta-\beta)^2+4\beta\delta}}{2\beta}.
 \end{aligned}
\end{equation}
Note that these parameters $b$ and $d$ already appeared in the construction of the boundary vectors involved in the matrix ansatz for the 
single species open ASEP \cite{Sandow94}. 

The boundary vectors $\bbra{\cW}$ and $\kket{\cV}$ are explicitly given by 
\begin{equation} \label{eq:2ASEP_WV}
 \bbra{\cW} = \sum_{n=0}^{+\infty} \frac{H_n(0,c)}{\sqrt{(q;q)_n}}\bbra{n} \quad \mbox{and} \quad 
 \kket{\cV} = \sum_{n=0}^{+\infty} \frac{H_n(b,d)}{\sqrt{(q;q)_n}}\kket{n},
\end{equation}
where the function $H_n(u,v)$ is defined in \eqref{eq:delH}.
It is indeed straightforward to check that they satisfy the relations \eqref{eq:boundaed} using the recurrence relation \eqref{eq:reccurence_H}.

\subsection{Computation of physical quantities}

Particle currents and densities are physical observable of prime interest because they can characterize phase transitions in the thermodynamic limit 
and they provide a good description of the macroscopic behavior of the system. 
The first step toward the computation of the mean currents and densities is the determination of the normalisation $\cZ_L$.

\subsubsection{Normalisation}

It will be particularly useful for following computations to study a slightly more general quantity defined as
\begin{equation}
 \cZ_L(\xi) = \bbra{\cW}(\cE+ \xi \cA + \cD)^L\kket{\cV}.
\end{equation}
The normalisation is easily recovered as $\cZ_L=\cZ_L(1)$. We can interpret $\cZ_L(\xi)$ as playing the role of a partition function 
with fugacity $\xi$ for particles of species $1$.
This generalized normalisation can be expressed exactly as \cite{caley}
\begin{equation} \label{eq:2ASEP_normalisation}
 \cZ_L(\xi) = \oint \frac{dz}{4\pi i z}w(z;r_0\xi) \, \Theta(z;0,c|r_0\xi) \, \Theta(z;b,d|r_0\xi) \, \left(\frac{(1+z)(1+z^{-1})}{1-q}\right)^L
\end{equation}
where
\begin{equation}
 w(z;\lambda)=\frac{(q,z^2,z^{-2};q)_{\infty}}{(\lambda z,\lambda z^{-1};q)_{\infty}}
\end{equation}
and 
\begin{equation}
 \Theta(z;u,v|\lambda) = \frac{(\lambda u,\lambda v;q)_{\infty}}{(uz,uz^{-1},vz,vz^{-1};q)_{\infty}} 
  {}_2 \phi_2\left[\genfracz{\lambda z,\lambda z^{-1}}{\lambda u, \lambda v} \Bigg| q, uv \right].
\end{equation}
The definitions of the $q$-Pochhammer $(\, \cdot \, ;q)_{\infty}$ and of the $q$-hypergeometric function ${}_2 \phi_2$ are recalled in appendix \ref{app:q_cal}.

We now introduce a quantity $\cZ_{L,k}$ that will play an important role in the multi-species generalization of the model. 
It will be in particular involved in the computation of the normalization and of the mean particle currents and densities. 
We define for all $0 \leq k \leq L$ 
\begin{equation} \label{eq:Z_Lk}
 \cZ_{L,k} = \cZ_L(\xi)\Big|_{\xi^k} = \bbra{\cW}(\cE+ \xi \cA + \cD)^L\kket{\cV}\Big|_{\xi^k}
\end{equation}
where for any polynomial, $p(t)\big|_{t^k}$ denotes the coefficient of $t^k$ in $p(t)$. $\cZ_{L,k}$ can hence be computed from 
\eqref{eq:2ASEP_normalisation} as 
\begin{equation}
 \cZ_{L,k} = \frac{1}{k!}\frac{\partial^k}{\partial \xi^k}\cZ_L(\xi) \Big|_{\xi=0}.
\end{equation}

In words $\cZ_{L,k}$ is the sum of the stationary weights of the configurations involving exactly $k$ particles of species $1$. 
Thus the probability to observe exactly $k$ particles of species $1$ on the lattice in the stationary state is given by
\begin{equation}\label{prob-k.part1}
P_k = \cZ_{L,k}/\cZ_L.
\end{equation}

\subsubsection{Particle currents}

Using the matrix product form of the stationary distribution \eqref{eq:2ASEP_Matrix_Ansatz} we can express the mean particle current of 
species $2$ as follows
\begin{equation} \label{eq:2ASEP_J2}
 J^{(2)} = \frac{1}{\cZ_L} \bbra{\cW}\cC^{k-1}(\cD\cE-q\cE\cD+\cD\cA-q\cA\cD)\cC^{L-k-1}\kket{\cV} = \frac{\cZ_{L-1}}{\cZ_L}
\end{equation}
where $\cC = \cE + \cA + \cD$.
The last equality in \eqref{eq:2ASEP_J2} is obtained using the commutation relations \eqref{eq:comEAD} of the matrix ansatz algebra.
The mean value of the current in the stationary state is constant all along the lattice and does not depend on the particular 
bond between sites $k$ and $k+1$ where it is computed.

We can also compute the mean particle current of species $1$ from the matrix product solution
\begin{equation} \label{eq:2ASEP_J1}
 J^{(1)} = \frac{1}{\cZ_L} \bbra{\cW}\cC^{k-1}(\cA\cE-q\cE\cA-\cD\cA+q\cA\cD)\cC^{L-k-1}\kket{\cV} = 0
\end{equation}
where the last equality is obtained using the commutation relations \eqref{eq:comEAD}. This result is consistent with the fact that particles
of species $1$ cannot enter nor leave the system at the right boundary and hence there cannot be a particle current in the stationary state.

Note that from expression \eqref{eq:2ASEP_J2} and from the explicit formula of the normalisation \eqref{eq:2ASEP_normalisation}
it is possible to derive the asymptotic behavior
of $J^{(2)}$ in the thermodynamic limit (\textit{i.e} for large system size $L$ going to infinity), and to draw the exact phase diagram. We do not 
study the thermodynamic limit in this paper, the reader is invited to look in \cite{caley} for details.

\subsubsection{Particle densities}

The average densities of particles of species $1$ and $2$ are also interesting quantities to describe the macroscopic behavior of the system in 
the thermodynamic limit. These quantities have been computed exactly for a finite size lattice in \cite{caley}. For the sake of simplicity,
we reproduce here only the result for the average density of species $1$.
Using the matrix product formalism it can be expressed as
\begin{equation} \label{eq:2ASEP_density1}
 \rho^{(1)} = \frac{1}{L} \frac{1}{\cZ_L} \sum_{k=1}^L \bbra{\cW}\cC^{k-1}\cA \, \cC^{L-k}\kket{\cV}
\end{equation}
It is straightforward to see that we have the equality
\begin{equation} \label{eq:2ASEP_density2}
 \rho^{(1)} = \frac{1}{L} \frac{\partial}{\partial \xi} \ln \cZ_L(\xi) \Big|_{\xi=1}
\end{equation}
which provides an explicit expression of the average density using the explicit formula \eqref{eq:2ASEP_normalisation}.

A direct computation also shows that $\rho^{(1)}$ is the mean number of particles of species $1$ present on the lattice in the stationary state 
(normalized by the lattice size $L$)
\begin{equation}\label{eq:mean1}
 \rho^{(1)} = \frac{1}{L} \sum_{k=0}^L k P_k
\end{equation}
where $P_k$ defined in \eqref{prob-k.part1} is the probability to observe $k$ particle of species $1$ on the lattice.

Note that from formula \eqref{eq:2ASEP_density2} (and from a similar formula for the average density $\rho^{(2)}$ of species $2$, see \cite{caley}),
it is possible to extract the asymptotic behavior in the limit $L$ going to infinity of the average densities $\rho^{(1)}$ and $\rho^{(2)}$. 
It provides important information about the macroscopic behavior of the system in the different phases.
The reader can refer to \cite{caley} for details.

\subsection{Totally asymmetric and symmetric limits}

We consider briefly here the totally asymmetric limit $q=0$ and the symmetric limit $q=1$ for which the algebraic structure of the 
matrix product solution is simpler and the computations are easier.

\subsubsection{Totally asymmetric case\label{sect:2TASEP}}

We first study the totally asymmetric case $q=0$. In order for the boundary rates to be consistent with the bulk dynamics, we need to have $\delta = 0$ and
$\tilde\gamma =0$. The latter leads us to two choices: either $\gamma = 0$ or $\gamma = 1-\alpha$. 

The first possibility gives a model
where species $1$ completely disappears in the stationary state (each configuration involving at least one particle of species
$1$ has a vanishing probability in the stationary state). The stationary state thus enjoys in this case the same properties as the usual single species
TASEP. 

The second possibility $\gamma = 1-\alpha$ provides a more interesting model for which all species of particle are present in the steady state with non-vanishing 
probabilities. For this model the algebraic relations satisfied by the matrices and boundary vectors simplify to
\begin{equation}\label{eq:TASEP_comEAD}
\begin{aligned}
 & \cD\cE= \cE+\cD, \\
 & \cA\cE = \cA, \\
 & \cD\cA = \cA,
\end{aligned}
\end{equation}
 and
 \begin{equation}\label{eq:TASEP_boundAED}
  \begin{cases}
   \bbra{\cW} \cE = \bbra{\cW} \\
   \bbra{\cW} \cA  = \frac{1-\alpha}{\alpha} \bbra{\cW}
  \end{cases}
 \quad \mbox{and} \quad  \cD \kket{\cV} = \frac{1}{\beta}\kket{\cV}.
 \end{equation}
It allows us to compute the normalisation $\cZ_L(\xi) = \bbra{\cW}(\cE + \xi \cA + \cD)^L\kket{\cV}$ by introducing
$\widetilde\cE = \cE + \xi \cA$. This operator satisfies the relations $\cD \widetilde\cE = \cD + \widetilde\cE$ and 
$\bbra{\cW} \widetilde\cE = \left(1+\xi\frac{1-\alpha}{\alpha}\right)\bbra{\cW}$, so that $\widetilde\cE$, $\cD$, $\bbra{\cW}$ and 
$\kket{\cV}$ fulfill the same algebraic relations as for the usual single species TASEP.
Using the results of \cite{DerridaEHP93} we thus deduce that
\begin{eqnarray} \label{eq:ZL_2TASEP}
 \cZ_L(\xi) & = &  \bbra{\cW}(\cE + \xi \cA + \cD)^L\kket{\cV} = \bbra{\cW}(\widetilde\cE + \cD)^L\kket{\cV} \\
 & = & \sum_{p=1}^L \frac{p(2L-1-p)!}{L!(L-p)!} 
 \frac{\left(1+\xi\frac{1-\alpha}{\alpha}\right)^{p+1}-\left(\frac{1}{\beta}\right)^{p+1}}{1+\xi\frac{1-\alpha}{\alpha}-\frac{1}{\beta}}.
\end{eqnarray}
It is then easy to deduce that
\begin{equation}\label{ZL-TASEP}
 \cZ_L = \sum_{p=1}^L \frac{p(2L-1-p)!}{L!(L-p)!} 
 \frac{\frac{1}{\alpha^{p+1}}-\frac{1}{\beta^{p+1}}}{\frac{1}{\alpha}-\frac{1}{\beta}}
\end{equation}
and
\begin{equation}
 \cZ_{L,k} =  \left(\frac{1-\alpha}{\alpha}\right)^k \sum_{p=k}^L \frac{p(2L-1-p)!}{L!(L-p)!} \sum_{j=k}^p 
 \begin{pmatrix}   j \\k    \end{pmatrix} 
 \frac{1}{\beta^{p-j}}.
\end{equation}

%
%

\subsubsection{Symmetric case}

We study here the symmetric case $q=1$. In this setting, the value of the boundary rate $\tilde\gamma$ is given by $\tilde\gamma = \gamma$.

For this model the algebraic relations satisfied by the matrices and boundary vectors are
\begin{equation}\label{eq:SSEP_comEAD}
\begin{aligned}
 & \cD\cE-\cE\cD = \cE+\cD, \\
 & \cA\cE-\cE\cA = \cA, \\
 & \cD\cA-\cA\cD = \cA,
\end{aligned}
\end{equation}
 
 \begin{equation}\label{eq:SSEP_boundAED}
  \begin{cases}
   \bbra{\cW} \cE = \frac{1}{\alpha+\gamma}\bbra{\cW} \\
   \bbra{\cW} \left(\gamma \cE - \alpha \cA + \gamma \cD \right) = 0
  \end{cases}
 \quad \mbox{and} \quad \left(\delta \cE - \beta \cD \right)\kket{\cV} = -\kket{\cV}
 \end{equation}
 It is straightforward to check that taking $\cA = r_0(\cE +\cD)$ with $r_0$ given in \eqref{def:EAD} is consistent with all these algebraic relations. The algebra thus reduces to the one 
 used to solve the single species SSEP. We can use the results for the single species SSEP \cite{DerrReview} to compute
\begin{eqnarray}
 \cZ_L(\xi) & = & \bbra{\cW}(\cE + \xi \cA + \cD)^L\kket{\cV} = (1+\xi r_0)^L \bbra{\cW}(\cE + \cD)^L\kket{\cV} \\
 & = & (1+\xi r_0)^L\frac{(\beta+\delta)^L}{\beta^L}  
 \frac{\Gamma\left(\frac{1}{\alpha+\gamma}+\frac{1}{\beta+\delta}+L\right)}{\Gamma\left(\frac{1}{\alpha+\gamma}+\frac{1}{\beta+\delta}\right)}
 \llangle \cW|\cV \rrangle,
\end{eqnarray}
where the gamma function satisfies $\Gamma(z+1)=z\Gamma(z)$.

A direct computation then yields
\begin{equation}\label{eq:ZLk}
 \cZ_{L,k} = r_0^k \begin{pmatrix}
                  L \\ k 
                 \end{pmatrix}
\frac{(\beta+\delta)^L}{\beta^L}  
 \frac{\Gamma\left(\frac{1}{\alpha+\gamma}+\frac{1}{\beta+\delta}+L\right)}{\Gamma\left(\frac{1}{\alpha+\gamma}+\frac{1}{\beta+\delta}\right)}
 \llangle \cW|\cV \rrangle
\end{equation}
and 
\begin{equation}
 P_k=\frac{\cZ_{L,k}}{\cZ_L} = \left(\frac{r_0}{1+r_0}\right)^k \left(\frac{1}{1+r_0}\right)^{L-k} \begin{pmatrix}
                                                                                             L \\ k 
                                                                                           \end{pmatrix},
\end{equation}
which is a binomial distribution with parameter $r_0/(1+r_0)= \gamma/(\alpha+\gamma)$ corresponding to the particle density of species $1$ in the left reservoir.

From this result we deduce the mean density for the species 1:
\begin{equation}\label{eq:rho1}
\rho^{(1)} = \frac{r_0}{1+r_0}.
\end{equation}
This confirms that the species 1 is at equilibrium: there is no current and the density in the left reservoir coincides with the mean density.

\section{3-species generalisation \label{sec:3ASEP}}

\subsection{Presentation of the model}
The 3-species ASEP is a straightforward generalisation of the 2-species model. It describes three species of particles (labelled 1, 2, 3) in addition to the vacancies (still labelled 0). The bulk interaction is very similar to the 2-species case. The new feature is contained in the interaction with the reservoirs, as can be seen in the
 boundary matrices
\begin{equation*}
    B = \begin{pmatrix}
        -\alpha - \gamma & 0 & 0 & 0
        \\
        \gamma & -\alpha & 0 & \tilde\gamma
        \\
        0 & 0 & 0 & 0
        \\
        \alpha & \alpha & 0 & -\tilde\gamma
    \end{pmatrix},
    \qquad
    \overline{B} = \begin{pmatrix}
        -\delta & 0 & 0 & \beta
        \\
        0 & -\delta_1  & \beta_1 & 0
        \\
        0 & \delta_1  & -\beta_1 & 0
        \\
        \delta & 0 & 0 &  -\beta
    \end{pmatrix},
\end{equation*}
where $\beta_1$ and $\delta_1$ are new parameters, independent from $\alpha, \beta, \gamma, \delta$. 

Remark that when these new parameters are free (in the sense $\beta_1\neq\beta$ and $\delta_1\neq\delta$), 
the boundary matrix $\overline B$ does not fall in the integrability class given in \cite{CrampeFRV16}. 
Moreover, there does not seem to exist a reflection matrix $\bar K(z)$ obeying the reflection equation and such that 
$\overline B= \frac12 \frac{d}{dz} \bar K(z)\Big|_{z=1}$. Hence, we are away from the integrability point.

The boundary and bulk matrices lead to the following stochastic rates
\begin{eqnarray}
\begin{array}{|c|c|c|} 
\hline 
\mbox{Left} & \mbox{Bulk} & \mbox{Right} 
\\ \hline 
\hspace{0.5cm} 0\, \xrightarrow{\ \gamma\ } \,1\hspace{0.5cm}&\hspace{0.5cm} 10\, \xrightarrow{\ 1\ } \,01\hspace{0.5cm},\hspace{0.5cm} 01\, \xrightarrow{\ q\ }\, 10\hspace{0.5cm}&\hspace{0.5cm} 0\, \xrightarrow{\ \delta\ } \,3\hspace{0.5cm}
\\
\hspace{0.5cm} 0\, \xrightarrow{\ \alpha\ } \,3\hspace{0.5cm}&\hspace{0.5cm} 20\, \xrightarrow{\ 1\ }\, 02\hspace{0.5cm},\hspace{0.5cm} 02\, \xrightarrow{\ q\ }\, 20\hspace{0.5cm}&\hspace{0.5cm} 3\, \xrightarrow{\ \beta\ } \,0\hspace{0.5cm}
\\
\hspace{0.5cm} 1\, \xrightarrow{\ \alpha\ } \,3\hspace{0.5cm}&\hspace{0.5cm} 30\, \xrightarrow{\ 1\ }\, 03\hspace{0.5cm},\hspace{0.5cm} 03\, \xrightarrow{\ q\ }\, 30\hspace{0.5cm}&\hspace{0.5cm} 2\, \xrightarrow{\ \beta_1\ } \,1\hspace{0.5cm}
\\
\hspace{0.5cm} 3\, \xrightarrow{\ \tilde\gamma\ } \,1\hspace{0.5cm}&\hspace{0.5cm} 21\, \xrightarrow{\ 1\ } \,12\hspace{0.5cm},\hspace{0.5cm} 12\, \xrightarrow{\ q\ }\, 21\hspace{0.5cm}& \hspace{0.5cm} 1\, \xrightarrow{\ \delta_1\ } \,2\hspace{0.5cm}
\\
&\hspace{0.5cm} 31\, \xrightarrow{\ 1\ } \,13\hspace{0.5cm},\hspace{0.5cm} 13\, \xrightarrow{\ q\ }\, 31\hspace{0.5cm}& 
\\
\hline
 \end{array}
\end{eqnarray}
Remark that all the species (including vacancies) interact non-trivialy with the reservoirs.

\subsection{Matrix product solution}

\subsubsection{Algebraic relations}

The stationary state $\steady$ is given in a matrix product form
\begin{equation} \label{eq:3ASEP_Matrix_Ansatz}
 \steady = \frac{1}{Z_L} \bbra{W} \begin{pmatrix}
                                       X_0 \\ X_1 \\ X_2 \\ X_3
                                      \end{pmatrix} \otimes
                                      \begin{pmatrix}
                                       X_0 \\ X_1 \\ X_2 \\ X_3
                                      \end{pmatrix} \otimes \dots \otimes
                                      \begin{pmatrix}
                                       X_0 \\ X_1 \\ X_2 \\ X_3
                                      \end{pmatrix} \kket{V},
\end{equation}
where the normalisation is expressed as $Z_L = \bbra{W}(X_0+X_1+X_2+X_3)^L \kket{V}$.

The matrices satisfy the following algebraic relations
\begin{equation}\label{eq:3ASEP_comX}
\begin{aligned}
 & X_3 X_0-qX_0 X_3 = X_0+X_3, & \\
 & X_2 X_0-qX_0 X_2 = 0, \quad &X_1 X_0-qX_0 X_1 = 0, \\
 & X_3 X_1-qX_1 X_3 = 0, \quad &X_2 X_1-qX_1 X_2 = 0, \\
 & X_3 X_2-qX_2 X_3 = 0. &
\end{aligned}
\end{equation}
 The relations on the boundary vectors are
 \begin{equation}\label{eq:3ASEP_boundX}
  \begin{cases}
   \bbra{W} X_0 = \frac{1}{\alpha+\gamma}\bbra{W} \\
   \bbra{W} \left(\alpha X_0 - \alpha X_1 + \tilde\gamma X_3 \right) = 0
  \end{cases}
 \quad \mbox{and} \quad 
  \begin{cases}
 \left(\delta X_0 - \beta X_3 \right)\kket{V} = \kket{V} \\
 \left(\delta_1 X_1 - \beta_1 X_2 \right)\kket{V} = 0
  \end{cases}
 \end{equation}
 
 These equations are equivalent to the telescopic relations
\begin{equation}
 m \begin{pmatrix}
   X_0 \\ X_1 \\ X_2 \\ X_3
   \end{pmatrix} \otimes
   \begin{pmatrix}
   X_0 \\ X_1 \\ X_2 \\ X_3
   \end{pmatrix} = 
   \begin{pmatrix}
   X_0 \\ X_1 \\ X_2 \\ X_3
   \end{pmatrix} \otimes
   \begin{pmatrix}
   -1 \\ 0 \\ 0 \\ 1
   \end{pmatrix}-
   \begin{pmatrix}
   -1 \\ 0 \\ 0 \\ 1
   \end{pmatrix} \otimes
   \begin{pmatrix}
   X_0 \\ X_1 \\ X_2 \\ X_3
   \end{pmatrix}
\end{equation}
and
\begin{equation}
 \bbra{W}B \begin{pmatrix}
               X_0 \\ X_1 \\ X_2 \\ X_3
               \end{pmatrix} =
 \bbra{W} \begin{pmatrix}
            -1 \\ 0 \\ 0 \\ 1
            \end{pmatrix}, \qquad 
 \overline{B} \begin{pmatrix}
                X_0 \\ X_1 \\ X_2 \\ X_3
                \end{pmatrix} \kket{V} =   
 -\begin{pmatrix}
  -1 \\ 0 \\ 0 \\ 1
  \end{pmatrix} \kket{V}
\end{equation}
which ensure that \eqref{eq:3ASEP_Matrix_Ansatz} provides a correct expression of the steady state.

\subsubsection{Representation}

The matrices $X_0$, $X_1$, $X_2$ and $X_3$ can be constructed as tensor product of the operators $\ope$, $\opd$ and $\opA$ as follows

 \begin{equation}\label{eq:MA3species}
\begin{array}{lll}
 X_0 &= \displaystyle\frac{1}{1-q}(\opid+\ope) \otimes \opid &= \cE \otimes \opid, \\[2ex]
 X_1 &=\displaystyle\frac{r_0}{1-q}\opA \otimes \opA &= \cA \otimes \opA, \\[2ex]
 X_2 &=\displaystyle\frac{r_0}{1-q}\opA \otimes \opd &= \cA \otimes \opd, \\[2ex]
 X_3 &=\displaystyle\frac{1}{1-q}(\opid+\opd) \otimes \opid &= \cD \otimes \opid.
\end{array}
\end{equation}
The second equalities in the construction \eqref{eq:MA3species} points out the factorization of the 2-species model
studied in the previous section. 
It will remains true in the general multi-species case studied in section \ref{sec:NASEP} where we will present
this factorization property more explicitly.
A similar factorization also happens for the boundary vectors
\begin{equation}
 \bbra{W} = \bbra{\cW} \otimes \bbra{0} \quad \mbox{and} \quad \kket{V} = \kket{\cV} \otimes \kket{v^{(1)}},
\end{equation}
where the vectors $\bbra{\cW}$ and $\kket{\cV}$ were defined in \eqref{eq:2ASEP_WV}.
The row vector $\bbra{0}$ is the first element of the Fock space basis and satisfies $\bbra{0} \opA = \bbra{0}$. The column vector 
$\kket{v^{(1)}}$ satisfies $\left(\delta_1 \opA - \beta_1 \opd \right)\kket{v^{(1)}}=0$. It is explicitly given by
\begin{equation}
 \kket{v^{(1)}} = \sum_{n=0}^{+\infty} \left(\frac{\delta_1}{\beta_1}\right)^n\frac{q^{n(n-1)/2}}{\sqrt{(q;q)_n}}\kket{n},
\end{equation}
so that $\llangle 0|v^{(1)}\rrangle =1$.

The matrix product structure of the stationary state is simple enough to perform computation of simple observables. We address this question in 
the following subsection.

\subsection{Computation of physical quantities}
\subsubsection{Partition function}
The first step toward the computation of physical quantities such as particle currents and densities is the computation of the normalisation 
$Z_L = \bbra{W} C^L \kket{V}$ where $C=X_0+X_1+X_2+X_3$.
Using the explicit representation, we start with the decomposition
$$
C=\sum_{j=0}^{3} X_j=\Big(\cE+\cD\Big)\otimes\opid+ \cA\otimes \big(\opA+\opd\big).
$$

This allows to compute 
$$
Z_L= \sum_{\ell=0}^L \bbra{0}\big(\opA+\opd\big)^\ell\kket{{v}^{(1)}} \ \cZ_{L,\ell},
$$
where $\cZ_{L,\ell}$ has been defined in \eqref{eq:Z_Lk}.

From the boundary conditions \eqref{eq:boundtilde} and the $q$-binomial development \eqref{eq:q-binomial}, it is easy to see that 
\begin{equation}
\bbra{0}\big(\opA+\opd\big)^L\kket{{v}^{(1)}}=\sum_{j=0}^L \left[\begin{array}{c} L\\ j\end{array}\right]_q\, (r_1)^j\,q^{j(j-1)/2}=(-r_1;q)_L,
\end{equation}
where $r_1=\frac{\delta_1}{\beta_1}$.
Then 
\begin{equation}
Z_L= \sum_{\ell=0}^L (-r_1;q)_\ell \ \cZ_{L,\ell}.
\end{equation}
Assuming $0 < q < 1$, the factor $(-r_1;q)_\ell$ increases with $\ell$, and approaches a limiting value
$(-r_1;q)_\infty$.  Thus one can see that with respect to the 2-species case, for which
$\displaystyle
\cZ_L= \sum_{\ell=0}^L  \cZ_{L,\ell}$, there is a reweighting of the stationary state in favour of
configurations with larger values of $\ell$.  That is, in favour of those configurations with a larger number
of particles of species 1 or 2.  We can equivalently consider it as a damping factor on configurations
containing a large number of particles of species 3 or vacancies.
It would be interesting to see how this damping affects the asymptotic behavior of the partition 
function $Z_L$ for large system size.

For the totally asymmetric case ($q = 0$) and the symmetric case ($q = 1$), the behaviour is somewhat
different.  We will discuss this in section~\ref{sec:NTASEPSSEP} where we compute $Z_L$ explicitly in these
limits.

\subsubsection{Particle currents}

We can now turn to the computation of the mean particle currents, that can be easily expressed using the matrix product structure. 
For the fastest species (species $3$) we have
\begin{equation} \label{eq:3ASEP_J3}
 J^{(3)} = \frac{1}{Z_L} \bbra{\cW}C^{k-1}\Big(X_3(X_0+X_1+X_2)-q(X_0+X_1+X_2)X_3\Big)C^{L-k-1}\kket{\cV} = \frac{Z_{L-1}}{Z_L},
\end{equation}
where the last equality is obtained using the commutation relations \eqref{eq:3ASEP_comX}. For the particles of species $1$ and $2$ we have
\begin{equation} \label{eq:3ASEP_J2}
 J^{(2)} = \frac{1}{Z_L} \bbra{\cW}C^{k-1}\Big(X_2(X_0+X_1)-q(X_0+X_1)X_2-X_3 X_2+qX_2 X_3\Big)C^{L-k-1}\kket{\cV} = 0
\end{equation}
and a similar computation yields $J^{(1)}=0$.

Note that one can also compute mean densities for each species: we postpone this calculation to the general case of $N$-species models, see next section.

\section{A N-species model \label{sec:NASEP}}
Now, we consider a $N$-species ASEP, labeling the species $1, 2, ... ,N$ and still labeling the vacancies $0$. 
Thus, the 2-species ASEP studied in section \ref{sec:2ASEP} corresponds to $N=2$ 
while the 3-species ASEP studied in section \ref{sec:3ASEP} corresponds to $N=3$.
In this section we will assume that $N\geq3$.

\subsection{Presentation of the model\label{ssec:modelNASEP}}

Let $\tau_i \in \{0,1,.., N\}$, and
$\bm\tau = (\tau_1, \ldots ,\tau_L)$ give the lattice configuration.  In the bulk, the hopping rates are
\begin{align*}
    & \tau_i \tau_{i+1} \xrightarrow{1} \tau_{i+1} \tau_i, \qquad \tau_i > \tau_{i+1},
    \\
    & \tau_i \tau_{i+1} \xrightarrow{q} \tau_{i+1} \tau_i, \qquad \tau_i < \tau_{i+1}.
\end{align*}
We will use two types of boundary matrices, $B$ for the left boundary and $\bar B$ for the right boundary. Their explicit form reads
\begin{equation}\label{eq:BBbar}
    B = \begin{pmatrix}
        -\alpha - \gamma &  0 & 0\ \cdots\ 0 & 0
        \\
        \gamma & -\alpha & 0\ \cdots\ 0 & \tilde\gamma
        \\
       \begin{array}{c} 0  \\ \vdots  \\ 0  \end{array} & \begin{array}{c} 0  \\ \vdots  \\ 0  \end{array}
        & \fbox{\rule{.5ex}{0cm}$\mathfrak{B}_{N-2}(\bar \alpha,\bar \gamma)$\rule[-2ex]{0cm}{6ex}\rule{.5ex}{0cm}} & \begin{array}{c} 0  \\ \vdots  \\ 0  \end{array}
        \\
        \alpha & \alpha & 0\ \cdots\ 0 & -\tilde\gamma
    \end{pmatrix},
    \quad
    \overline{B} = \begin{pmatrix}
        -\delta & 0\ \cdots\ 0 & \beta
        \\
         \begin{array}{c} 0  \\ \vdots  \\ 0  \end{array}
          &\fbox{\rule{.5ex}{0cm}$\mathfrak{B}_{N-1}(\bar \delta,\bar \beta)$\rule[-2ex]{0cm}{6ex}\rule{.5ex}{0cm}}  &  \begin{array}{c} 0  \\ \vdots  \\ 0  \end{array}
        \\
        \delta & 0\ \cdots\ 0 &  -\beta
    \end{pmatrix},
\end{equation}
where  $\bar \alpha=\{\alpha_2,\alpha_4,...\}$, $\bar \gamma=\{\gamma_2,\gamma_4,...\}$, $\bar \beta=\{\beta_1,\beta_3,...\}$
and $\bar \delta=\{\delta_1,\delta_3,...\}$.

The form of the sub-matrices $\mathfrak{B}_{M}(\bar x,\bar y)$ ($M=N-2$ or $N-1$) depends on the parity of $M$:
\begin{equation*}
\mathfrak{B}_{2m+1} (\bar x,\bar y)= \begin{pmatrix}
       -x_{1} & 0 & 0 & 0 & 0 & 0 &  y_{1}
        \\
        0 & \ddots & 0 & 0 & 0 & \iddots & 0
        \\
        0 & 0 & -x_{m} & 0 & y_{m} & 0 & 0
        \\
        0 & 0 & 0 & 0 & 0 & 0 & 0 
        \\
        0 & 0 & x_{m} & 0 & -y_{m} & 0 & 0
        \\
        0 & \iddots & 0 & 0 & 0 & \ddots & 0
        \\
        x_{1} & 0 & 0 & 0 & 0 & 0 & -y_{1}
    \end{pmatrix},
\end{equation*}
\begin{equation*}
\mathfrak{B}_{2m}(\bar x,\bar y) = \begin{pmatrix}
        -x_{1} & 0 & 0 & 0 & 0 & y_{1}
        \\
        0 & \ddots  & 0 & 0 & \iddots & 0
        \\
        0 & 0 & -x_{m} & y_{m} & 0 & 0
        \\
        0 & 0 & x_{m}  & -y_{m} & 0 & 0
        \\
        0 & \iddots  & 0 & 0 & \ddots & 0
        \\
        x_{1} & 0 & 0 & 0 & 0 & -y_{1}
    \end{pmatrix}.
\end{equation*}
Remark that when $N$ is odd, $\fB_{2m+1}(\bar x,\bar y)$ occurs in the left boundary matrix $B$, with 
$(\bar x,\bar y)=(\bar \alpha,\bar \gamma)$, the last $\alpha_{2j}$ being $\alpha_{N-3}$, and $\fB_{2m}(\bar x,\bar y)$ 
occurs in the right boundary matrix $\overline B$, 
with $(\bar x,\bar y)=(\bar \delta,\bar \beta)$, the last $\beta_{2j-1}$ being $\beta_{N-2}$. 

When $N$ is even, one exchanges $\fB_{2m+1}$ and $\fB_{2m}$ keeping $(\bar \alpha,\bar \gamma)$ on the left boundary 
(then the last $\alpha_{2j}$ is $\alpha_{N-2}$), and $(\bar \delta,\bar \beta)$ on the right boundary 
(then the last $\beta_{2j-1}$ is $\beta_{N-3}$).

Note that the boundaries are integrable when all parameters $\alpha_{j}$ (resp. $\beta_{j}$, $\gamma_{j}$, $\delta_{j}$) are equal to $\alpha$ (resp. $\beta$,
$\gamma$, $\delta$) \cite{CrampeFRV16}.  In the following, we keep the parameters $\alpha_{j}$, $\beta_{j}$, $\gamma_{j}$ and $\delta_{j}$ free. 

The boundary matrices translate into local exchange rules on the boundaries (the corresponding probability rates are indicated above the 
arrows)
\begin{equation}
\begin{array}{|ccc|ccc|}
\hline
 & \text{Left} & & & \text{Right} &
\\ \hline
     0 & \xrightarrow{\gamma} & 1 & 0 & \xrightarrow{\delta} & N
    \\
     0 & \xrightarrow{\alpha} & N & N & \xrightarrow{\beta} & 0
    \\
     1 & \xrightarrow{\alpha} & N &   &  &                   
    \\
     i & \xrightarrow{\alpha_{2i-2}} & N+1-i & i & \xrightarrow{\delta_{2i-1}} & N-i 
    \\
    N+1-i & \xrightarrow{\gamma_{2i-2}} & i & N-i & \xrightarrow{\beta_{2i-1}} & i 
    \\
     & 2 \leq i \leq \lfloor\frac{N}{2}\rfloor & & & 1 \leq i \leq \lfloor\frac{N-1}{2}\rfloor & 
    \\ \hline
\end{array}
\end{equation}
Remark that the combination of left and right boundaries mixes all species $1\leq i\leq N-1$.

\subsection{Stationary state in matrix product form}

We can construct a matrix product form for the stationary state in a way similar to \cite{Uchiyama08}.  
To start, we introduce the algebra $\fX(N)$, whose generators
$X_i$, $i = 0, \ldots, N$ obey the following algebraic relations
\begin{equation}\label{com:X}
X_i X_j - q X_j X_i = x_i X_j - x_j X_i, \quad i > j,
\end{equation}
 with scalars
\begin{equation*}
    x_0 = -1, \qquad x_j = 0, \quad j=1,...,N-1, \qquad x_N = 1.
\end{equation*}
Then  the stationary probabilities can be written as
\begin{equation}\label{eq:proba}
    \cS(\bm\tau) = \frac{1}{Z_L}\bbra{W}X_{\tau_1} \ldots X_{\tau_L}\kket{V},
\end{equation}
with
\begin{equation*}
    Z_L = \bbra{W} C^L \kket{V}, \qquad C = \sum_{j=0}^{N}X_j.
\end{equation*}
The relations on the boundary vectors $\bbra{W}$ and $\kket{V}$ are
\begin{eqnarray}
&&\label{eq:W}
\begin{aligned}
    & \bbra{W}\Big[(\alpha + \gamma)X_0+x_0 \Big]=0,
    \\
    & \bbra{W}\Big[\alpha X_0 + \alpha X_1  -\tilde \gamma X_N - x_N\Big] = 0,
    \\
    & \bbra{W}\Big[\alpha_{2i-2}X_i  -  \gamma_{2i-2}X_{N+1-i}\Big] = 0,\quad 2\leq i \leq \left\lfloor\frac{N}{2}\right\rfloor,
\end{aligned}
\end{eqnarray}
\begin{eqnarray}
&&\label{eq:V}
\begin{aligned}
    & \Big[\delta X_0-\beta X_N-x_0\Big] \kket{V} = 0,
    \\
    & \Big[\delta_{2i-1} X_i - \beta_{2i-1} X_{N-i}\Big] \kket{V} = 0,\quad 1\leq i \leq \left\lfloor\frac{N-1}{2}\right\rfloor,
\end{aligned}
\end{eqnarray}
where the parameters $\alpha$, $\alpha_{j}$, $\beta$,... are the ones entering the boundary matrices $B$ and $\overline{B}$, see \eqref{eq:BBbar}.

The existence of non-trivial matrices $X_i$ generating the algebra $\fX(N)$ and of vectors $\bbra{W}$ and $\kket{V}$ is ensured 
by the construction of an explicit representation in which relations \eqref{com:X}, \eqref{eq:W} and
\eqref{eq:V} are fulfilled. 

The commutation relations \eqref{com:X} are equivalent to the telescopic relations
\begin{equation}
 m \begin{pmatrix}
   X_0 \\ X_1 \\ \vdots \\ X_{N-1} \\ X_N
   \end{pmatrix} \otimes
   \begin{pmatrix}
   X_0 \\ X_1 \\ \vdots \\ X_{N-1} \\ X_N
   \end{pmatrix} = 
   \begin{pmatrix}
   X_0 \\ X_1 \\ \vdots \\ X_{N-1} \\ X_N
   \end{pmatrix} \otimes
   \begin{pmatrix}
   -1 \\ 0 \\ \vdots \\ 0 \\ 1
   \end{pmatrix}-
   \begin{pmatrix}
   -1 \\ 0 \\ \vdots \\ 0 \\ 1
   \end{pmatrix} \otimes
   \begin{pmatrix}
   X_0 \\ X_1 \\ \vdots \\ X_{N-1} \\ X_N
   \end{pmatrix}
\end{equation}
and the relations on the boundary vectors \eqref{eq:W} and \eqref{eq:V} can be also rewritten as
\begin{equation}
 \bbra{W}B \begin{pmatrix}
               X_0 \\ X_1 \\ \vdots \\ X_{N-1} \\ X_N
               \end{pmatrix} =
 \bbra{W} \begin{pmatrix}
           -1 \\ 0 \\ \vdots \\ 0 \\ 1
            \end{pmatrix}, \qquad 
 \overline{B} \begin{pmatrix}
                X_0 \\ X_1 \\ \vdots \\ X_{N-1} \\ X_N
                \end{pmatrix} \kket{V} =   
 -\begin{pmatrix}
  -1 \\ 0 \\ \vdots \\ 0 \\ 1
  \end{pmatrix} \kket{V}
\end{equation}
which ensure that \eqref{eq:proba} provides a correct expression of the steady state.

\subsection{Representation for the Matrix product algebra}
\subsubsection{Factorisation of the 2-ASEP algebra}

One can separate  two types of species in writing:
\begin{equation}
\begin{aligned}
    X_0 & =   \cE \otimes \opid, \qquad
    &X_N  =   \cD \otimes \opid,
 \\
   X_i &= \cA\otimes \mathring{X}_i,\qquad &1\leq i\leq N-1,
\end{aligned}
\end{equation}
where the operators $\cE$, $\cD$ and $\cA$ have been introduced in section \ref{sec:2ASEP}.
Then, from the exchange relations \eqref{eq:comEAD} and \eqref{com:X} one deduces
\begin{equation}\label{eq:Xeq}
\mathring{X}_i \mathring{X}_j - q \mathring{X}_j \mathring{X}_i = 0, \quad i > j.
\end{equation}
The relation \eqref{eq:Xeq} defines an algebra $\mathring{\fX}(N)$ generated by $\{\mathring{X}_i,\, i=1,...N-1\}$ which corresponds to  
a thermodynamical equilibrium. It already appeared in \cite{Ari} to study multi-species ASEP model with reflexive boundaries.
However, the representation used in \cite{Ari} is not compatible with the present boundary conditions.

Writing 
\begin{equation*}
    \bbra{W} = \bbra{\cW}\otimes\bbra{\mathring{W}}\quad\mbox{and} \quad \kket{V} = \kket{\cV}\otimes\kket{\mathring{V}},
\end{equation*}
the boundary relations \eqref{eq:W}, \eqref{eq:V} factorise as well.
We recover the conditions \eqref{eq:boundAED} for the vectors $\bbra{\cW}$ and $\kket{\cV}$. Thus the representation used in section \ref{sec:2ASEP} can still be used for these vectors. 
For the vectors $\bbra{\mathring{W}}$ and $\kket{\mathring{V}}$, we obtain
\begin{equation}\label{eq:boundeq}
\begin{aligned}
    & \bbra{\mathring{W}} \mathring{X}_1 = \bbra{\mathring{W}} 
    \\
    &\bbra{\mathring{W}} \Big[\alpha_{2i-2}\, \mathring{X}_i  -  \gamma_{2i-2}\, \mathring{X}_{N+1-i}\Big] = 0,\quad  2\leq i \leq \left\lfloor\frac{N}{2}\right\rfloor,
    \\
    & \Big[\delta_{2i-1}\mathring{X}_i-\beta_{2i-1}\mathring{X}_{N-i}\Big] \kket{\mathring{V}}=0,\quad  1\leq
    i \leq \left\lfloor\frac{N-1}{2}\right\rfloor.
\end{aligned}
\end{equation}
The next section presents an explicit representation for these vectors.

\subsubsection{Realization of the algebra $\mathring{\fX}(N)$ and its boundary vectors}
The algebra $\mathring{\fX}(N)$ generated by $\mathring{X}_1$, $\mathring{X}_2$,..., $\mathring{X}_{N-1}$ as well as the vectors 
$ \bbra{\overline{\cW}}$ and $\kket{\overline{\cV}}$ can be constructed as a tensor product of $N-2$ algebras $(\opA,\ope,\opd)$: 
\begin{equation}\label{X:equil}
\begin{array}{cclllllllllll}
    \mathring{X}_1  &= &\opA &\otimes &\opid &\otimes &\opid &\otimes &\opid&\otimes &  \opid  &\otimes &\\
    \mathring{X}_2  &= &\opd &\otimes &\ope &\otimes &\opA&\otimes &\opid&\otimes &  \opid  &\otimes &\\
    \mathring{X}_3  &= &\opd &\otimes &\ope &\otimes &\opd&\otimes &\ope &\otimes & \opA &\otimes&
    \\
    \mathring{X}_4  &= &\opd &\otimes &\ope &\otimes &\opd&\otimes &\ope &\otimes & \opd &\otimes&
    \\
    \vdots && \vdots& &\vdots & &\vdots & & \vdots & & \vdots & &\dots
    \\
    \mathring{X}_{N-3}  &= &\opd &\otimes &\ope &\otimes &\opd&\otimes &\ope &\otimes & \opd &\otimes&
    \\
    \mathring{X}_{N-2}  &= &\opd &\otimes &\ope &\otimes &\opd&\otimes &\opA &\otimes & \opid &\otimes&
    \\
    \mathring{X}_{N-1}  &= &\opd &\otimes &\opA &\otimes &\opid &\otimes &\opid &\otimes &\opid &\otimes&
\end{array}
\end{equation}
In words, the tensor products organize into successive columns, that are alternatively either 
$\left(\begin{array}{c} \opA\\ \opd\\ \vdots \\ \opd\end{array}\right)$ or 
$\left(\begin{array}{c}\ope\\ \vdots \\ \ope \\ \opA\end{array}\right)$. 
The size of the columns decreases by 1 for each tensor product, and they are completed by the identity $\opid$ to keep a full size of $N-1$,
with the rule that on the right of $\opA$ or $\opid$, there is always $\opid$.

This construction is related to two recursion procedures that build $\mathring{\fX}(N)$ from $\mathring{\fX}(N-1)$. 
Indeed, denoting by $\mathring{Y}_j$ the generators of $\mathring{\fX}(N-1)$, one can check that
\begin{equation}\label{embed1}
 \mathring{X}_1  =\opA \otimes \opid \quad\mbox{and}\quad  \mathring{X}_{j+1}  =\opd \otimes \mathring{Y}_{j}\,,\ j=1,2,...N-2
 \end{equation}
obey the relations \eqref{eq:Xeq} for $\mathring{\fX}(N)$. The same is true for
\begin{equation}\label{embed2}
  \mathring{X}_{j}  =\ope \otimes \mathring{Y}_{j}\,,\ j=1,2,...N-2 \quad\mbox{and}\quad  \mathring{X}_{N-1}  =\opA \otimes \opid .
 \end{equation}
 These two embeddings allow to construct by recursion several representations of $\mathring{\fX}(N)$.
The representation \eqref{X:equil} results from an alternating choice of these two possible embeddings, starting
from \eqref{embed1} for $\mathring{\fX}(N)$ and going down recursively to $\mathring{\fX}(2)$, which is trivial. 
This choice is dictated by the boundary conditions we have chosen.

Indeed, using the representation \eqref{X:equil} for the $\mathring{\fX}(N)$ algebras, one can see that
the boundary vectors $\bbra{\mathring{W}}$ and $\kket{\mathring{V}}$ can be constructed as
\begin{eqnarray}\label{eq:WV1}
\bbra{\mathring{W}} &=& \bbra{0} \otimes \bbra{{w}^{(2)}} \otimes \bbra{0} \otimes \bbra{{w}^{(4)}} \otimes \dots
\\
\kket{\mathring{V}} &=& \kket{{v}^{(1)}} \otimes \kket{0} \otimes \kket{{v}^{(3)}} \otimes \kket{0} \otimes \dots
\label{eq:WV2}
\end{eqnarray}
where 
\begin{equation}\label{eq:boundtilde}
\begin{array}{lll}
\opd \kket{{v}^{(2j-1)}}= r_{2j-1} \opA\kket{{v}^{(2j-1)}} &;\quad \opA \kket{0} = \kket{0}&;\quad r_{2j-1}=\frac{\delta_{2j-1}}{\beta_{2j-1}}
\\[1.2ex]
\bbra{0} \opA =  \bbra{0}&;\quad  \bbra{{w}^{(2j)}}\ope = r_{2j}\,  \bbra{{w}^{(2j)}} \opA &;\quad r_{2j}=\frac{\gamma_{2j}}{\alpha_{2j}}
\end{array}
\quad \,j=1,2,.....
\end{equation}
Note that the last $r_k$ is $r_{N-2}$, in accordance with the counting done in section \ref{ssec:modelNASEP}.
We remind that $r_0=\frac{\gamma}{\alpha}$ (see section \ref{sec:2ASEP}).

Similarly to the 3-species case, one can obtain the following explicit expression for the boundary vectors:
\begin{equation}
 \bbra{w^{(2j)}} = \sum_{n=0}^{+\infty} \left(r_{2j}\right)^n\,\frac{q^{n(n-1)/2}}{\sqrt{(q;q)_n}}\,\bbra{n}\quad \mbox{and}\quad
 \kket{v^{(2j-1)}} = \sum_{n=0}^{+\infty} \left(r_{2j-1}\right)^n\,\frac{q^{n(n-1)/2}}{\sqrt{(q;q)_n}}\,\kket{n}.
\end{equation}
It  shows that $\langle\!\langle{\mathring{W}}|{\mathring{V}}\rangle\!\rangle =1$.

To see that \eqref{eq:WV1}, \eqref{eq:WV2} and \eqref{eq:boundtilde} fulfill relations \eqref{eq:boundeq}, 
one has to detail the 'positions' of the operators $\ope$, $\opd$, $\opA$ in the representation \eqref{X:equil} of the generators $\mathring{X}_{j}$
(i.e. in which space of the tensor product they occur). Apart from $\mathring{X}_{1}$, whose operator content is rather clear, one gets the following operator content for the generators $\mathring{X}_{j}$:
\begin{eqnarray*}
&&\text{For }\ 2\leq j\leq\left\lfloor\frac{N}2\right\rfloor\ :\
\begin{cases} \opd \text{ occurs in positions } 2k-1\ \text{for}\ 1\leq k\leq j-1 \\
\ope \text{ occurs in positions } 2k\ \text{for}\  1\leq k\leq j-1 \\
\opA \text{ occurs in position } 2j-1 \end{cases}
\\
&&\text{For }\ \left\lfloor\frac{N}2\right\rfloor< j\leq N-1\ :\
\begin{cases} \opd \text{ occurs in positions } 2(N-k)-1\ \text{for}\ j\leq k\leq N-1 \\
\ope \text{ occurs in positions } 2(N-k)\ \text{for}\ j< k\leq N-1 \\
\opA \text{ occurs in position } 2(N-j) \end{cases}
\end{eqnarray*}
If the 'position' of $\opA$ is greater than the total number of spaces (i.e. $N-2$), then this operator does not occur in $\mathring{X}_j$.
With these rules, it is easy to check \eqref{eq:boundeq} from \eqref{eq:WV1}, \eqref{eq:WV2} and \eqref{eq:boundtilde}.

%
%
%

\subsection{Recursive construction of the partition function}

We start with the decomposition
$$
C=\sum_{j=0}^{N} X_j=\Big(\cE+\cD\Big)\otimes\opid+ \cA\otimes \mathring{C}
\quad \mbox{where} \quad \mathring{C}= \sum_{j=1}^{N-1} \mathring{X}_j.
$$

This allows to compute 
\begin{equation}\label{norm-N}
Z_L= \sum_{\ell=0}^L \mathring{Z}_\ell \ \cZ_{L,\ell}
\quad\mbox{with}\quad 
\mathring{Z}_\ell=\bbra{\mathring{W}}\,( \mathring{C})^\ell\, \kket{\mathring{V}}
\end{equation}
$\cZ_{L,\ell}$ has been computed in section \ref{sec:2ASEP}, see \eqref{eq:Z_Lk}.
As for the 3-species case, we have a damping factor on the number of particles of the fastest species (species
$N$) with respect to the 2-species case.

$\mathring{Z}_L$ can be computed through a recursion on the number of species:
\begin{eqnarray}
\mathring{Z}_L &=& \sum_{0\leq k_{N-2}\leq...\leq  k_{1}\leq L} \begin{bmatrix}L\\L-k_1,\ k_1-k_2,\ ...,\ k_{N-3}-k_{N-2},\ k_{N-2}\end{bmatrix}_q
\,q^{\fb(k_1,...,k_{N-2})}
\nonumber\\
&&\qquad\qquad\qquad\times \prod_{j=1}^{N-2} r_j^{k_j},
\label{eq:ZL-Nspecies}\\
\fb(k_1,...,k_{N-2})&=&\frac12\Big({k_1(k_1-1)+k_2(k_2-1)+...+k_{N-2}(k_{N-2}-1)}\Big)
\end{eqnarray}
where the $q$-deformed multinomial coefficient has been defined in \eqref{eq:q-multi}.

\medskip

\prf
We introduce the generators
$\{\mathring{Y}_1,...,\mathring{Y}_{N-2}\}$ for the algebra $\mathring{\fX}(N-1)$. Then, we define
$\mathring{C}(\mathring{Y})= \sum_{j=1}^{N-2} \mathring{Y}_j$ and $\mathring{Z}_L(\mathring{Y})$ the corresponding partition function 
depending on the boundary parameters $r_2,r_3,...,r_{N-2}$, while the original quantities will be noted $\mathring{C}(\mathring{X})$ and 
$\mathring{Z}_L(\mathring{X})$. Then from
$\mathring{C}(\mathring{X})=\opA\otimes\opid+\opd\otimes \mathring{C}(\mathring{Y})$ one deduces
\begin{equation}
\mathring{Z}_L(\mathring{X}) = \sum_{\ell=0}^L \Big(\bbra{0}\big(\opA+\xi\opd\big)^L\kket{{v}^{(1)}}\Big)\Big|_{\xi^\ell}\, \mathring{Z}_\ell(\mathring{Y}).
\end{equation}

From the boundary conditions \eqref{eq:boundtilde} and the $q$-binomial development \eqref{eq:q-binomial}, it is easy to see that 
\begin{equation}
\bbra{0}\big(\opA+\xi\opd\big)^L\kket{{v}^{(1)}}=\sum_{j=0}^L \left[\begin{array}{c} L\\ j\end{array}\right]_q\, (r_1\xi)^j\,q^{j(j-1)/2}=(-r_1\xi;q)_L.
\end{equation}
Then 
\begin{equation}
\mathring{Z}_L(\mathring{X}) = \sum_{\ell=0}^L \left[\begin{array}{c} L\\ \ell\end{array}\right]_q\, (r_1)^\ell\,q^{\ell(\ell-1)/2}\, \mathring{Z}_\ell(\mathring{Y}).
\label{eq:ZLstep1}
\end{equation}
To perform the recursion, one also needs 
\begin{eqnarray}
\mathring{Z}_L(\mathring{Y}) &=& \sum_{\ell=0}^L \Big(\bbra{{w}^{(2)}}\big(\opA+\xi\ope\big)^L\kket{0}\Big)\Big|_{\xi^\ell}\, \mathring{Z}_\ell(\mathring{Y}'),
\\
&=&
\sum_{\ell=0}^L \left[\begin{array}{c} L\\ \ell\end{array}\right]_q\, (r_2)^\ell\,q^{\ell(\ell-1)/2}\, \mathring{Z}_\ell(\mathring{Y}'),
\label{eq:ZLstep2}
\end{eqnarray}
where now $\mathring{Y}'_j$ generates the algebra $\mathring{\fX}(N-2)$.

Using alternatively formulas \eqref{eq:ZLstep1} and \eqref{eq:ZLstep2}, one finally obtains \eqref{eq:ZL-Nspecies}.

\subsection{Physical quantities}
\subsubsection{Particle currents}
Using the bulk relations of the algebra \eqref{com:X}, we can compute the current of particles of species $N$:
\begin{equation*}
    J^{(N)} = \frac{1}{Z_L} \bbra{W} C^{k-1}\left(\sum_{i=0}^{N-1}( X_N X_i - q X_i X_N)\right) C^{L-k-1} \kket{V}
            = \frac{Z_{L-1}}{Z_L}.
\end{equation*}
In the same way we find that the current associated to equilibrium species vanishes:
\begin{equation*}
\begin{aligned}
    J^{(j)} &= \frac{1}{Z_L} \bbra{W} 
    C^{k-1}\left(\sum_{i=0}^{j-1}( X_{j} X_i - q X_i X_{j})+\sum_{i=j+1}^{N}( q X_{j} X_i - X_i X_{j})\right) C^{L-k-1} \kket{V}\\
            &= 0 \quad \mbox{for}\quad 1 \leq j \leq N-1.
\end{aligned}
\end{equation*}

\subsubsection{Particle densities}
We would like to compute the average density of each species $i$:
\begin{equation*}
    \rho^{(i)} = \frac{1}{L}\frac{1}{Z_L}\sum_{k=1}^L \bbra{W}C^{k-1} X_i C^{L-k}\kket{V}.
\end{equation*}
A common way to compute this quantity is to introduce the parameters
$\bar\zeta=\{\zeta_1, \ldots, \zeta_N\}$ and to define 
\begin{equation*}
    Z_L(\bar\zeta)\equiv  \bbra{W} \big(X_0 + \zeta_1 X_1 + \ldots + \zeta_N X_N\big)^L \kket{V},
\end{equation*}
which plays the role of a grand canonical partition function.  The average density is then expressed easily as
\begin{equation*}
    \rho^{(i)}
    =
    \frac{1}{L} \frac{\partial}{\partial \zeta_i}
        \log Z_L(\bar\zeta)\big|_{\zeta_1 = \ldots = \zeta_N = 1}.
\end{equation*}
The calculation of $Z_L(\bar\zeta)$ can be done recursively, following the same steps as for $Z_L$. We have
\begin{eqnarray}
C(\bar\zeta) &=& X_0+\sum_{j=1}^{N} \zeta_j X_j 
= \Big(\cE+\zeta_N\cD\Big)\otimes\opid +\cA\otimes \mathring{C}(\zeta_1,...,\zeta_{N-1})
\\
Z_L(\bar\zeta)&=& \sum_{\ell=0}^L \mathring{Z}_\ell(\zeta_1,...,\zeta_{N-1}) \ \Big(\cZ_L(\xi,\zeta_N)\Big)\!\Big|_{\xi^\ell}
\end{eqnarray}
where we have introduced 
$$
\begin{cases} 
\mathring{Z}_\ell(\zeta_1,...,\zeta_{N-1})=\bbra{\mathring{W}}\,\big( \mathring{C}(\zeta_1,...,\zeta_{N-1})\big)^\ell\, \kket{\mathring{V}},
\\ 
\cZ_L(\xi,\zeta_N)=\bbra{\cW}\big(\cE+\zeta_N\cD+\xi\,\cA\big)^L\kket{\cV}. 
\end{cases}
$$
Similarly to the computation of $\mathring{Z}_L$, a recursion yields
\begin{eqnarray}\label{ZequilN}
\mathring{Z}_L(\zeta_1,...,\zeta_{N-1}) &=& \sum_{0\leq k_{N-2}\leq...\leq  k_{1}\leq L}
\begin{bmatrix}L\\L-k_1,\ k_1-k_2,\ ...,\ k_{N-3}-k_{N-2},\ k_{N-2}\end{bmatrix}_q
\,q^{\fb(k_1,...,k_{N-2})}
\nonumber\\
&\times& \Big(\prod_{j=1}^{N-2}r_j^{k_j}\Big)\,\zeta_1^{L-k_{1}}\,\zeta_2^{k_{1}-k_{2}}\,\zeta_3^{k_{2}-k_{3}}\,...\zeta_{N-1}^{k_{N-2}},\\
\fb(k_1,...,k_{N-2})&=&\frac12\Big({k_1(k_1-1)+k_2(k_2-1)+...+k_{N-2}(k_{N-2}-1)}\Big).
\end{eqnarray}

Finally, 
we remark that for $1\leq k \leq N-1$ we have 
$$
\zeta_k\frac{\partial}{\partial\zeta_k}\mathring{Z}_L(\zeta_1,...,\zeta_{N-1})=\Big(r_{k-1}\frac{\partial}{\partial r_{k-1}}-r_{k}\frac{\partial}{\partial r_{k}}\Big)
\mathring{Z}_L(\zeta_1,...,\zeta_{N-1}),
$$ 
so that the average density of equilibrium species can be obtained directly from the non-deformed partition function $Z_L$
\begin{eqnarray}\label{eq:density}
\rho^{(k)}  &=& \frac{1}{L} \Big(r_{k-1}\frac{\partial}{\partial r_{k-1}}-r_{k}\frac{\partial}{\partial r_{k}}\Big) \log Z_L, \qquad 1\leq k \leq N-1
\end{eqnarray}
where $Z_L$ is given by \eqref{norm-N}, $r_0=\gamma/\alpha$ and by convention $r_{N-1}=0$.

\subsection{Totally asymmetric and symmetric limits}
\label{sec:NTASEPSSEP}

We consider again the totally asymmetric limit $q=0$ and the symmetric limit $q=1$ for which the partition function 
and the mean densities can be computed more explicitly.

\subsubsection{Totally asymmetric case}
Due to the factor $q^{\fb(k_1,...,k_{N-2})}$ in \eqref{ZequilN}, the limit $q\to0$ imposes  $k_j\in\{0,1\}$, $\forall j$. Then, we get
\begin{equation}
\mathring{Z}_l(\zeta_1,...,\zeta_{N-1}) = 
\begin{cases} 1, & l = 0,  \\
        \zeta_1^{l-1}\sum_{p=0}^{N-2} \zeta_{p+1}\prod_{j=1}^p r_j, & l > 0.
\end{cases}
\end{equation}

Note that for $l > 0$, $$\mathring{Z}_l = \mathring{Z}_l(\zeta_1 = 1,...,\zeta_{N-1} = 1) =1+\sum_{p=1}^{N-2}\prod_{j=1}^p r_j.$$   
Thus we see a limited
damping effect in the totally asymmetric case: configurations with no equilibrium particles are disfavoured
(they are reweighted by $\mathring{Z}_0 = 1$),
but all other configurations receive an equal boost ($\mathring{Z}_l > 1$, $l > 0$).

Using \eqref{ZL-TASEP}, it implies the following expression for the 
normalization
\begin{eqnarray}
 Z_L(\zeta_1,...,\zeta_{N})\Big|_{\zeta_N=1} &=& \left(1+\sum_{k=1}^{N-2}\frac{\zeta_{k+1}}{\zeta_1}\prod_{j=1}^{k}r_j\right)
 \cZ_L(\zeta_1)-\left(\sum_{k=1}^{N-2}\frac{\zeta_{k+1}}{\zeta_1}\prod_{j=1}^{k}r_j\right)
 \cZ_L(0),
\nonu
\mbox{where} \quad \cZ_L(\xi)&=& \sum_{p=1}^L \frac{p(2L-1-p)!}{L!(L-p)!} 
 \frac{\big(\xi\frac{1-\alpha}\alpha+1\big)^{p+1}-\frac{1}{\beta^{p+1}}}{\xi\frac{1-\alpha}\alpha+1-\frac{1}{\beta}} \llangle \cW|\cV \rrangle
\end{eqnarray}
is just the normalisation for the 2-TASEP \eqref{eq:ZL_2TASEP}.

From this expression, one can deduce the expression of the mean density for equilibrium species. For such a purpose, 
we  introduce
\begin{equation}
 \cR = \sum_{k=1}^{N-2}\prod_{j=1}^{k}r_j \quad\text{ and }\quad \fz_L=\frac{\cZ_L(0)}{\cZ_L(1)}.
\end{equation}
Then
\begin{eqnarray}
 \rho^{(1)} &=&\frac{1+\cR}{1+(1-\fz_L)\cR}\ \rho^{(1)}_{2spec} -\frac{1}{L}\, \frac{(1-\fz_L)\cR}{1+(1-\fz_L)\cR},  \\
 \rho^{(p)} &=& \frac{1}{L}\, \frac{1-\fz_L}{1+(1-\fz_L)\cR}\,\prod_{j=1}^{p-1}r_j, \qquad \mbox{for} \quad 1< p \leq N-1,
\end{eqnarray}
where $$ \rho^{(1)}_{2spec}=\frac1L\frac{ \cZ_L'(1)}{ \cZ_L(1)}$$ is the density of species 1 in the 
2-TASEP\footnote{We remind that for the TASEP, one imposes $\gamma=1-\alpha$, see section \ref{sect:2TASEP}.}. 

Remark that we have
\begin{equation}
\sum_{p=1}^{N-1} \rho^{(p)} =\frac{1+\cR}{1+(1-\fz_L)\cR}\ \rho^{(1)}_{2spec},
\end{equation}
showing the effect of the number of equilibrium species on the mean density.

\subsubsection{Symmetric case}

In the limit $q\to 1$, the $q$-deformed multinomial coefficients become usual multinomial coefficients. A direct computation 
using Newton's multinomial theorem yields
\begin{equation}
 \mathring{Z}_\ell(\zeta_1,...,\zeta_{N-1}) = \left(\sum_{p=1}^{N-1} \zeta_p \prod_{j=1}^{p-1}r_j\right)^\ell.
\end{equation}
Using \eqref{eq:ZLk}, it implies the following expression for the normalization 
\begin{equation}
 Z_{L}(\zeta_1,...,\zeta_{N})\Big|_{\zeta_N=1} = \left(1+\sum_{p=1}^{N-1}\zeta_p\prod_{j=0}^{p-1}r_j\right)^L
\frac{(\beta+\delta)^L}{\beta^L}  
 \frac{\Gamma\left(\frac{1}{\alpha+\gamma}+\frac{1}{\beta+\delta}+L\right)}{\Gamma\left(\frac{1}{\alpha+\gamma}+\frac{1}{\beta+\delta}\right)}
 \llangle \cW|\cV \rrangle.
\end{equation}
Then, we can deduce the explicit expression of the mean density of equilibrium species
\begin{equation}
 \rho^{(k)} = \frac{\prod_{j=0}^{k-1} r_j}{1+\sum_{p=1}^{N-1}\prod_{j=0}^{p-1}r_j}, \qquad \mbox{for} \quad 1 \leq k \leq N-1.
\end{equation}

Remark that 
\begin{equation}
\sum_{k=1}^{N-1} \rho^{(k)}=\frac{R}{1+R} \mb{with} R=r_0\Big(1+\sum_{p=2}^{N-1}\prod_{j=1}^{p-1}r_j\Big)
\end{equation}
that has to be compared with the 2-ASEP result \eqref{eq:rho1}: one sees that $r_0$ has been replaced by $R>r_0$.
Note also that when $R$ becomes large, the summed density of the equilibrium species approaches $1$.  Thus we
see that adding equilibrium species induces the damping of the non-equilibrium species (species $N$ and vacancies, species $0$).

\section{Conclusion}

We presented in this paper a class of N-species ASEP with open boundary conditions. The injection and extraction rates of the particles of 
different species at the boundaries are rich enough to ensure the presence of all species in the stationary state. These boundary conditions
do not fall directly in the classes of integrable boundaries provided in \cite{CrampeFRV16} but appear as a generalization of them, the integrable case 
being recovered when the parameters are fixed to some specific values. Nevertheless, we expressed exactly the associated stationary state in 
a matrix product form using several tensor copies of a $q$-oscillator algebra. This point out the fact that a simple matrix product solution 
does not imply necessarily that the model is integrable, at least in the sense of the existence (based on Yang-Baxter and reflection equations) of commuting transfer matrices generating the Markov matrix.
Note that the reverse statement (\textit{i.e} Yang-Baxter/reflection integrability implies the existence of a simple matrix product solution)
has been argued to be true \cite{CrampeRV14}.

It would be very interesting to construct in matrix product form the stationary states associated to all the other classes of 
integrable boundaries provided in \cite{CrampeFRV16}. The 2-species case has now been settled \cite{Uchiyama08,CrampeMRV15,CrampeEMRV16,caley}
but we are still lacking for a solution for the general N-species case. In particular it would be nice to construct the stationary state
of the model pointed out in \cite{CantiniGDGW16} in a matrix product form because of its connection with Koornwinder polynomials.

\section*{Acknowledgments}
M.V. warmly thanks the LAPTh, where most of the work has been done, for hospitality and financial support. M.V. also acknowledges the 
financial support by the ERC under the Advanced Grant 694544 OMNES.

\appendix
\section{The $q$-calculus \label{app:q_cal}}
We introduce here notations and properties that will be needed all along the paper.
We first define the $q$-Pochhammer symbol
\begin{equation}
 (a_1,\dots,a_r;q)_n = \prod_{l=1}^{r} (a_l;q)_n
\end{equation}
where
\begin{equation}
 (a;q)_n = \prod_{k=0}^{n-1} (1-a q^k)
\end{equation}
Note that if $|q|<1$, these expressions also make sense when $n \rightarrow \infty$.   
It allows us to introduce the $q$-deformed binomial coefficient
\begin{equation}
 \begin{bmatrix}
  n \\ k
 \end{bmatrix}_q =
 \frac{(q;q)_n}{(q;q)_k (q;q)_{n-k}}.
\end{equation}
More generally, we will also use the $q$-deformed multinomial coefficient
\begin{equation}\label{eq:q-multi}
 \begin{bmatrix}
  n \\ k_1,\ k_2,\ ...,\ k_p
 \end{bmatrix}_q =
 \frac{(q;q)_n}{(q;q)_{k_1}(q;q)_{k_2}\cdots(q;q)_{k_p} (q;q)_{n-k_1-...-k_p}}.
\end{equation}
The $q$-deformed binomial coefficient satisfies a $q$-deformed triangle relation
\begin{equation}
 \begin{bmatrix}
  n+1 \\ k
 \end{bmatrix}_q = 
 \begin{bmatrix}
  n \\ k-1
 \end{bmatrix}_q +
 q^k\begin{bmatrix}
  n \\ k
 \end{bmatrix}_q.
\end{equation}
This allows us to prove by induction the two following properties that will be needed later. The $q$-Pochhammer expansion
\begin{equation}
 (a;q)_n = \sum_{k=0}^n  \begin{bmatrix}
  n \\ k
 \end{bmatrix}_q \, (-a)^k q^{k(k-1)/2}
\end{equation}
and the $q$-binomial development
\begin{equation}\label{eq:q-binomial}
(\mathbf{x}+\mathbf{y})^n=\sum_{k=0}^n  \begin{bmatrix}
  n \\ k
 \end{bmatrix}_q \,\mathbf{x}^{n-k} \mathbf{y}^{k},
\end{equation}
where $\mathbf{x}$ and $\mathbf{y}$ are non-commuting variables satisfying $\mathbf{yx}=q\mathbf{xy}$.

When dealing with the representation theory of matrix ansatz algebras, we will make use of the following polynomials
\begin{equation}\label{eq:delH}
    H_n(u,v) = \sum_{k=0}^n \begin{bmatrix}
  n \\ k
 \end{bmatrix}_q u^{n-k}v^k,
\end{equation}
which satisfies the recurrence relation
\begin{equation} \label{eq:reccurence_H}
    H_{n+1}(u,v)
    + (1 - q^n) u v H_{n-1}(u,v)
    = (u + v)H_n(u,v),
\end{equation}
with $H_{-1} = 0$ and $H_0 = 1$.

We will also need to define the $q$-hypergeometric functions
\begin{equation}
 {}_r \phi_s\left[\genfracz{a_1,\dots,a_r}{b_1,\dots,b_s} \Bigg| q, z \right] =
 \sum_{k=0}^{\infty} \frac{(a_1,\dots,a_r;q)_k}{(q,b_1,\dots,b_s;q)_k}\left((-1)^k q^{\left(\genfracz{k}{2}\right)}\right)^{1+s-r}z^k.
\end{equation}


\begin{thebibliography}{99}

\bibitem{Ari} C. Arita, 
\textsl{Remarks on the multi-species exclusion process with reflective boundaries},
J. Phys. \textbf{A 45} (2012) 155001 and \texttt{arXiv:1112.5585}.

\bibitem{AAMP1} C. Arita, A. Ayyer, K. Mallick and S. Prolhac,
\textsl{Recursive structures in the multi-species TASEP},
J. Phys. \textbf{A 44} (2011) 335004 and \texttt{arXiv:1104.3752}.

\bibitem{AAMP2} C. Arita, A. Ayyer, K. Mallick and S. Prolhac, 
\textsl{Generalized matrix Ansatz in the multi-species exclusion process - the partially asymmetric case}, 
J. Phys. \textbf{A 45} (2012) 195001 and \texttt{arXiv:1201.0388}.

\bibitem{arita} C. Arita, K. Mallick,
\textsl{Matrix product solution to an inhomogeneous multi-species TASEP},
J. Phys. A: Math. Theor. 46 085002 (2013) 
\texttt{arXiv:1209.1913}.

\bibitem{caley} 
Arvind Ayyer, Caley Finn, Dipankar Roy, 
\textsl{Matrix product solution of a left-permeable two-species asymmetric exclusion process},
\texttt{arXiv:1708.09153}.

\bibitem{ALS} A. Ayyer, J.L. Lebowitz and E.R. Speer,
\textsl{On the Two Species Asymmetric Exclusion Process with Semi-Permeable Boundaries},
J. Stat. Phys. \textbf{135} (2009) 1009 and \texttt{arXiv:0807.2423}.

\bibitem{AyyerR16}
A. Ayyer and D. {Roy},
\textsl{The exact phase diagram for a class of multispecies asymmetric exclusion processes},
\texttt{arXiv:1611.01943}.

\bibitem{BorodinFPS07}
A.~Borodin, P.~L. Ferrari, M.~Pr{\"a}hofer, and T.~Sasamoto, ``{Fluctuation
  properties of the TASEP with periodic initial configuration},'' {\em Journal
  of Statistical Physics} {\bfseries 129} no.~5, (2007) 1055--1080.

\bibitem{Cantini15}
L.~Cantini, ``{Asymmetric Simple Exclusion Process with open boundaries and
  {K}oornwinder polynomials},''
  \href{http://arxiv.org/abs/1506.00284}{{\ttfamily arXiv:1506.00284
  [math-ph]}}.

\bibitem{CantiniGDGW16}
L.~Cantini, A.~Garbali, J.~de~Gier, and M.~Wheeler, ``{Koornwinder polynomials
  and the stationary multi-species asymmetric exclusion process with open
  boundaries},'' \href{http://dx.doi.org/10.1088/1751-8113/49/44/444002}{{\em
  J. Phys. A: Math. Theor.} {\bfseries 49} no.~44, (2016) 444002},
  \href{http://arxiv.org/abs/1607.00039}{{\ttfamily arXiv:1607.00039
  [math-ph]}}.
  
\bibitem{CantiniDGW15}
L.~Cantini, J.~de~Gier, and M.~Wheeler, ``{Matrix product formula for
  {M}acdonald polynomials},''
  \href{http://dx.doi.org/10.1088/1751-8113/48/38/384001}{{\em J. Phys. A:
  Math. Theor.} {\bfseries 48} no.~38, (2015) 384001},
  \href{http://arxiv.org/abs/1505.00287}{{\ttfamily arXiv:1505.00287
  [math-ph]}}.
  
\bibitem{ChouMZ11}
T.~Chou, K.~Mallick, and R.~Zia, ``{Non-equilibrium statistical mechanics: from
  a paradigmatic model to biological transport},''
  \href{http://dx.doi.org/10.1088/0034-4885/74/11/116601}{{\em Reports on
  progress in physics} {\bfseries 74} no.~11, (2011) 116601},
  \href{http://arxiv.org/abs/1110.1783}{{\ttfamily arXiv:1110.1783
  [cond-mat.stat-mech]}}.

\bibitem{CrampeEMRV16}
N.~Crampe, M.~Evans, K.~Mallick, E.~Ragoucy, and M.~Vanicat, ``{Matrix product
  solution to a 2-species {TASEP} with open integrable boundaries},''
  \href{http://dx.doi.org/10.1088/1751-8113/49/47/475001}{{\em J. Phys. A:
  Math. Theor.} {\bfseries 49} (2016) 475001},
  \href{http://arxiv.org/abs/1606.08148}{{\ttfamily arXiv:1606.08148
  [cond-mat.stat-mech]}}.

\bibitem{CrampeFRV16}
N.~Crampe, C.~Finn, E.~Ragoucy, and M.~Vanicat, ``{Integrable boundary
  conditions for multi-species {ASEP}},''
  \href{http://dx.doi.org/10.1088/1751-8113/49/37/375201}{{\em J. Phys. A:
  Math. Theor.} {\bfseries 49} no.~37, (2016) 375201},
  \href{http://arxiv.org/abs/1606.01018}{{\ttfamily arXiv:1606.01018
  [math-ph]}}.
  
\bibitem{CrampeMRV15}
N.~Crampe, K.~Mallick, E.~Ragoucy, and M.~Vanicat, ``{Open two-species
  exclusion processes with integrable boundaries},''
  \href{http://dx.doi.org/10.1088/1751-8113/48/17/175002}{{\em J. Phys. A:
  Math. Theor.} {\bfseries 48} no.~17, (2015) 175002},
  \href{http://arxiv.org/abs/1412.5939}{{\ttfamily arXiv:1412.5939
  [cond-mat.stat-mech]}}.

\bibitem{CrampeRV14}
N.~Crampe, E.~Ragoucy, and M.~Vanicat, ``{Integrable approach to simple
  exclusion processes with boundaries. Review and progress},''
  \href{http://dx.doi.org/10.1088/1742-5468/2014/11/P11032}{{\em J. Stat.
  Mech.} (2014) P11032}, \href{http://arxiv.org/abs/1408.5357}{{\ttfamily
  arXiv:1408.5357 [math-ph]}}.
  
\bibitem{Derrida98}
B.~Derrida, ``{An exactly soluble non-equilibrium system: the asymmetric simple
  exclusion process},''
  \href{http://dx.doi.org/10.1016/S0370-1573(98)00006-4}{{\em Physics Reports}
  {\bfseries 301} no.~1, (1998) 65--83}.
  
\bibitem{DerrReview} B. Derrida, 
\textsl{Non-equilibrium steady  states: fluctuations and large deviations of the density and of the current}, 
J. Stat. Mech. (2007) P07023 and \texttt{arXiv:cond-mat/0703762}.

\bibitem{DerridaEHP93}
B.~Derrida, M.~R. Evans, V.~Hakim, and V.~Pasquier, ``{Exact solution of a 1{D}
  asymmetric exclusion model using a matrix formulation},''
  \href{http://dx.doi.org/10.1088/0305-4470/26/7/011}{{\em Journal of Physics
  A: Mathematical and General} {\bfseries 26} no.~7, (1993) 1493}.
  
\bibitem{EvansFM09}
M.~Evans, P.~Ferrari, and K.~Mallick, ``{Matrix Representation of the
  Stationary Measure for the multi-species TASEP},''
  \href{http://dx.doi.org/10.1007/s10955-009-9696-2}{{\em J Stat Phys}
  {\bfseries 135} (2009) 217}, \href{http://arxiv.org/abs/0807.0327}{{\ttfamily
  arXiv:0807.0327 [math.PR]}}.
  
\bibitem{Ferrari91}
P.~A. Ferrari, ``{Microscopic shocks in one dimensional driven systems},'' in
  {\em {Annales de l'IHP Physique th{\'e}orique}}, vol.~55, pp.~637--655.
\newblock 1991.

\bibitem{FerrariKS91}
P.~Ferrari, C.~Kipnis, E.~Saada, {\em et~al.}, ``{Microscopic structure of
  travelling waves in the asymmetric simple exclusion process},'' {\em The
  Annals of Probability} {\bfseries 19} no.~1, (1991) 226--244.
  
\bibitem{FinnV17} C.~Finn and M.~Vanicat, ``{Matrix product construction for Koornwinder polynomials
and fluctuations of the current in the open ASEP},''
{\em J. Stat. Mech.} (2017) P023102, 
{\ttfamily arXiv:1610.08320 [math-ph]}

\bibitem{Krug91}
J.~Krug, ``{Boundary-induced phase transitions in driven diffusive systems},''
  {\em Physical review letters} {\bfseries 67} no.~14, (1991) 1882.
 
\bibitem{KunibaMO15}
A.~Kuniba, S.~Maruyama, and M.~Okado, ``{Multispecies TASEP and combinatorial
  $R^*$},'' \href{http://dx.doi.org/10.1088/1751-8113/48/34/34FT02}{{\em J.
  Phys. A: Math. Theor.} {\bfseries 48} (2015) 34FT02},
  \href{http://arxiv.org/abs/1506.04490}{{\ttfamily arXiv:1506.04490
  [math-ph]}}.

\bibitem{KunibaMO16}
A.~Kuniba, S.~Maruyama, and M.~Okado, ``{Multispecies TASEP and the tetrahedron
  equation},'' \href{http://dx.doi.org/10.1088/1751-8113/49/11/114001}{{\em J.
  Phys. A: Math. Theor.} {\bfseries 49} (2016) 114001},
  \href{http://arxiv.org/abs/1509.09018}{{\ttfamily arXiv:1509.09018
  [nlin.SI]}}.

\bibitem{Liggett85}
T.~M. Liggett, {\em {Interacting Particle Systems}}, vol.~276.
\newblock Springer-Verlag, New York, 1985.

\bibitem{Mandelshtam15}
O.~Mandelshtam, ``{Matrix ansatz and combinatorics of the $ k $-species
  PASEP},'' {\em arXiv preprint arXiv:1508.04115} (2015) .

\bibitem{Sandow94}
S.~Sandow, ``{Partially asymmetric exclusion process with open boundaries},''
  \href{http://dx.doi.org/10.1103/PhysRevE.50.2660}{{\em Phys. Rev. E}
  {\bfseries 50} no.~4, (1994) 2660}.

\bibitem{SchuetzD93}
G.~Sch{\"u}tz and E.~Domany, ``Phase transitions in an exactly soluble
  one-dimensional exclusion process,''
  \href{http://dx.doi.org/10.1007/BF01048050}{{\em J. Stat. Phys.} {\bfseries
  72} no.~1-2, (1993) 277--296},
  \href{http://arxiv.org/abs/cond-mat/9303038}{{\ttfamily
  arXiv:cond-mat/9303038}}.

\bibitem{Spitzer70}
F.~Spitzer, ``{Interaction of {M}arkov processes},''
  \href{http://dx.doi.org/10.1016/0001-8708(70)90034-4}{{\em Advances in
  Mathematics} {\bfseries 5} no.~2, (1970) 246--290}.

\bibitem{TracyW09}
C.~A. Tracy and H.~Widom, ``{Asymptotics in ASEP with step initial
  condition},'' {\em Communications in Mathematical Physics} {\bfseries 290}
  no.~1, (2009) 129--154.  

\bibitem{Uchiyama08}
M. Uchiyama,
\textsl{Two-species asymmetric simple exclusion process with open boundaries},
Chaos, Solitons \& Fractals \textbf{35} (2008) 398 - 407.

\bibitem{UchiyamaSW04}
M.~Uchiyama, T.~Sasamoto, and M.~Wadati, ``{Asymmetric simple exclusion process
  with open boundaries and Askey--Wilson polynomials},'' {\em Journal of
  Physics A: Mathematical and General} {\bfseries 37} no.~18, (2004) 4985.

\end{thebibliography}
\end{document}